\documentclass[prd,twocolumn,floatfix,showpacs,preprintnumbers,nofootinbib, superscriptaddress]{revtex4}
\usepackage[utf8x]{inputenc}
\usepackage{mathrsfs}
\usepackage{amssymb}
\usepackage{amsmath}
\usepackage{graphicx}

\usepackage[caption = false]{subfig}
\usepackage[normalem]{ulem}
\usepackage{xcolor}
\definecolor{lcolor}{rgb}{0.5,0,0}
\definecolor{citcolor}{rgb}{0,0.3,0.0}

\usepackage[breaklinks,colorlinks,urlcolor=blue,citecolor=citcolor,linkcolor=lcolor]{hyperref}
\usepackage{mciteplus}

\newcommand{\rt}{{\mathbf{r}}}
\newcommand{\xt}{{\mathbf{x}}}
\newcommand{\bt}{{\mathbf{b}}}
\newcommand{\bti}{{\mathbf{b}_{i}}}
\newcommand{\yt}{{\mathbf{y}}}
\newcommand{\zt}{{\mathbf{z}}}

\newcommand{\Deltat}{{\boldsymbol{\Delta}}}

\newcommand{\tr}{\, \mathrm{Tr} \, }

\newcommand{\nc}{{N_\mathrm{c}}}

\newcommand{\gev}{\ \textrm{GeV}}
\newcommand{\fm}{\ \textrm{fm}}

\newcommand{\qs}{Q_\mathrm{s}}

\newcommand{\lqcd}{\Lambda_{\mathrm{QCD}}}
\newcommand{\as}{\alpha_{\mathrm{s}}}

\newcommand{\xbj}{{x}}

\newcommand{\xpom}{{x_\mathbb{P}}}

\newcommand{\der}{\mathrm{d}}

\newcommand{\A}{{\mathcal{A}}}

\begin{document}

\author{Heikki Mäntysaari}
\affiliation{
Department of Physics, University of Jyväskylä, %
 P.O. Box 35, 40014 University of Jyväskylä, Finland
}
\affiliation{
Helsinki Institute of Physics, P.O. Box 64, 00014 University of Helsinki, Finland
}

\author{Bj\"orn Schenke}
\affiliation{
Physics Department, Brookhaven National Laboratory, Upton, NY 11973, USA
}

\title{
Accessing the gluonic structure of light nuclei at the Electron Ion Collider
}

\pacs{24.85.+p, 13.60.-r}
% 24.85.+p: Quarks, gluons, and QCD in nuclear reactions
% 13.60.-r: Photon and charged-lepton interactions with hadrons

\preprint{}

\begin{abstract}
We show how exclusive vector meson production off light ions can be used to probe the spatial distribution of small-$x$ gluons in the deuteron and $^3$He wave functions. In particular, we demonstrate how short range repulsive nucleon-nucleon interactions affect the predicted coherent $J/\Psi$ production spectra. Fluctuations of the nucleon substructure are shown to have a significant effect on the incoherent cross section above $|t|\gtrsim 0.2\gev^2$. By explicitly performing the  JIMWLK evolution, we predict the $x$-dependence of coherent and incoherent cross sections in the EIC energy range.  Besides the increase of the average size of the nucleus with decreasing $x$, both the growth of the nucleons and subnucleonic hot spots are visible in the cross sections. The decreasing length scale of color charge fluctuations with decreasing $x$ is also present, but may not be observable for $|t|<1\gev^2$, if subnucleonic spatial fluctuations are present.
\end{abstract}

\maketitle

%%%%%%%%%%%%%%%%%%%%%%%%%
\section{Introduction}
%%%%%%%%%%%%%%%%%%%%%%%%%%

Electron-proton and electron-nucleus collisions can be used to precisely probe the internal structure and dynamics of protons and nuclei. Deep inelastic scattering (DIS) measurements of electrons on protons, performed at HERA, in which the electron emits a virtual photon which scatters off the target proton,  have provided a detailed picture of the internal quark and gluon structure of the proton~\cite{Aaron:2009aa,Abramowicz:2015mha}. These measurements have revealed that at high energies (small longitudinal momentum fraction $x$), the proton structure is dominated by gluons.

Recently, the authors have argued that the spatial distribution of small-$x$ gluons in the proton fluctuates event-by-event. This is evident from studying exclusive vector meson production. In coherent scattering where the target proton remains intact, the average shape of the proton is probed. In incoherent diffraction where the target dissociates, on the other hand, one is sensitive to the amount of event-by-event fluctuations~\cite{Mantysaari:2016ykx,Mantysaari:2016jaz} (see also~\cite{Miettinen:1978jb,Frankfurt:1993qi,Frankfurt:2008vi,Caldwell:2009ke,Lappi:2010dd}). These nucleon shape fluctuations have also been suggested to have a measurable effect in heavy nuclei~\cite{Mantysaari:2017dwh}.In addition to DIS experiments, the nucleon shape can be studied in hadronic collisions. In particular, the proton-lead collisions at the Large Hadron Collider (LHC) have revealed unexpected collective phenomena (for a review, see e.g.~\cite{Dusling:2015gta}). One potential source of the observed collectivity is the final state response to the initial state geometry. To verify this interpretation and to disentangle it from other sources of correlations, a good understanding of the proton geometry (and nucleon geometry in the nucleus) is required. It was shown that the flow measurements in LHC proton-lead collisions are compatible with a hydrodynamically evolving Quark Gluon Plasma (QGP), initiated with a Color Glass Condensate (CGC) initial condition, only if proton geometry fluctuations are included~\cite{Schenke:2014zha,Mantysaari:2017cni}. Similarly, the geometry fluctuations in the nucleons were found to be important in a global analysis of lead-lead and proton-lead flow data~\cite{Moreland:2018gsh}. Also, analysis of the elastic proton-proton differential cross section, measured by the TOTEM Collaboration~\cite{Antchev:2011zz} at high energy, has revealed indications for a hot spot structure of the proton~\cite{Albacete:2016pmp}.

The structure of light nuclei, such as the deuteron and helium at large $x$, is well known from low-energy scattering experiments. However, little is known about the spatial distribution of small-$x$ gluons in these systems.
To access the fundamental information on the small $x$ gluon structure in light nuclei, as well as to provide input to models aimed at describing deuteron-gold and helium-gold collisions at the Relativistic Heavy Ion Collider (RHIC) \cite{Abelev:2008ab,Aidala:2017ajz,PHENIX:2018lia}, one needs new measurements possible at a future electron ion collider (EIC) \,\cite{Accardi:2012qut}.

In this work we study the EIC's potential to constrain the spatial distribution of small-$x$ gluons in light nuclei via measurements of exclusive vector meson production. In particular, we compute differential coherent and incoherent cross sections for $J/\Psi$ production within both the IPsat model~\cite{Kowalski:2003hm} and the Color Glass Condensate framework, which includes the explicit solution of the JIMWLK~\cite{JalilianMarian:1996xn,JalilianMarian:1997jx,JalilianMarian:1997gr,Iancu:2001md} evolution equations.

This article is structure as follows. In Sec.~\ref{sec:nucstruct} we discuss how the deuteron and helium structure is obtained in terms of the nucleon constituents. In Sec.~\ref{sec:diffraction} it is shown how diffractive scattering processes are calculated at high energy in the dipole picture. The required dipole-nucleus scattering amplitudes encoding the target structure are obtained as discussed in Sec.~\ref{sec:dipoletarget}. The resulting energy evolution for the structure of light nuclei is illustrated in Sec.~\ref{sec:evolution}, and predictions for the future Electron Ion Collider are shown in Sec.~\ref{sec:predictions}. Our conclusions and outlook are presented in Sec.~\ref{sec:conclusions}.

\section{Structure of light ions}\label{sec:nucstruct}
The density profiles of light nuclei can be calculated rather accurately using theoretical methods to describe strongly correlated quantum systems. We will extract the nucleon positions in light ($A=2$, $3$) nuclei from such calculations configuration-by-configuration. In the high energy scattering processes at the EIC one is sensitive to the small-$x$ gluon distribution, about which we have to make additional assumptions given the distribution of nucleons. Future experimental data in combination with our calculations will be able to better constrain the small-$x$ structure of light nuclei.

In order to quantify the uncertainty in the current understanding of the deuteron structure, and to study the capabilities of the EIC for constraining the model uncertainties, we apply two different deuteron wave functions in this work. First, we use the deuteron wave function from \cite{Wiringa:1994wb,ArgonneWFweb}, obtained using the Argonne v18 (AV18) two-nucleon potential, including both $S$ and $P$ wave contributions, referred to as \emph{Argonne} in this manuscript.  This ab initio calculation includes both attractive and repulsive nucleon-nucleon interactions.

For comparison, we also employ the commonly used Hulthen wave function~\cite{Miller:2007ri} in which the distribution of the proton-neutron separation $d_\text{pn}$ is parametrized as
\begin{equation}
\label{eqref:hutlhen}
\phi_\text{pn}(d_\text{pn}) = \frac{1}{\sqrt{2\pi}} \frac{\sqrt{ab(a+b)}}{b-a} \frac{ e^{-a d_\text{pn}} - e^{-bd_\text{pn}}}{d_\text{pn}}.
\end{equation}
The experimentally determined parameters used in this work are $a=0.228\fm^{-1}$ and $b=1.18\fm^{-1}$~\cite{Miller:2007ri}. Note that these parameters are fixed by low energy data, and there is no \emph{a priori} reason why the small-$x$ gluonic distribution should resemble this precisely.

\begin{figure}[tb]
\centering
		\includegraphics[width=0.5\textwidth]{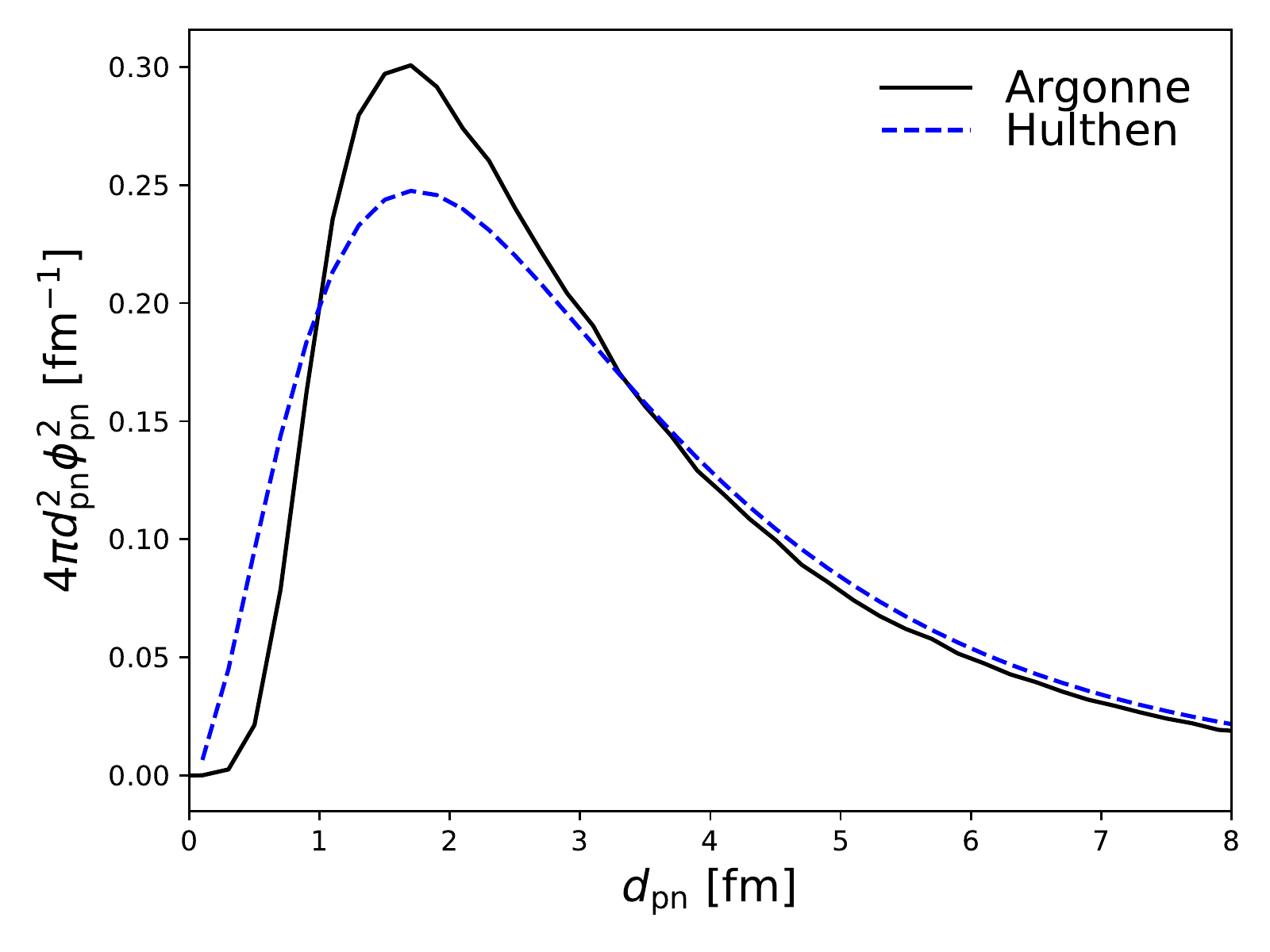} 
				\caption{Deuteron size distribution from the Hulthen and Argonne potential wave functions. }
		\label{fig:deuteron_size}
\end{figure}

The distribution of the proton-neutron separation $d_\mathrm{pn}$ (in 3 dimensions) obtained from these two parametrizations is shown in Fig.~\ref{fig:deuteron_size}. Both the Hulthen and Argonne potential wave functions produce deuterons with roughly the same root mean square size ($3.93$ fm in case of the Argonne potential, $4.07$ fm in case of Hulthen). The largest difference between the wave functions is that the repulsive short-range nucleon-nucleon interactions suppress small proton-neutron separations in the Argonne wave function compared to the Hulthen case. We note that recently the short range correlations in deuterons (and other light nuclei) have been studied in detail, in connection with the EMC effect, see e.g.~\cite{Weinstein:2010rt,Hen:2013oha,Hen:2016kwk}.

%\subsection{Helium 3}

The nucleonic structure of ${}^3$He is obtained from the Monte Carlo calculation with AV18+UIX interaction~\cite{Carlson:1997qn}. AV18 refers to the same two-nucleon potential used to obtain the deuteron configurations discussed above. In practice, we use the same database of configurations that is used in  Ref.~\cite{Nagle:2013lja}, and available in the PHOBOS Monte Carlo Glauber implementation~\cite{Loizides:2014vua}. These configurations reproduce the charge radii and form factors of Helium-3, as well as the relative separation of proton pairs.

\section{Diffractive deep inelastic scattering}
\label{sec:diffraction}
Diffractive vector meson production is a powerful probe of the spatial structure of nuclei, as the total momentum transfer, which is the Fourier conjugate to the impact parameter, is measurable. 
These processes are divided in two categories: in coherent scattering the target hadron remains in its ground state, and in incoherent diffraction the target dissociates. In either case there is no net color charge transferred to the probe, which leads to an experimentally observable rapidity gap between the produced vector meson and the target or target remnants.

In the Good-Walker picture~\cite{Good:1960ba}, the coherent cross section is obtained by calculating the average scattering amplitude for the Fock states of the probing virtual photon that diagonalize the scattering matrix. At high energy, these states are quark-antiquark states with fixed transverse separation $\rt$ scattering off a fixed target configuration. The coherent cross section  can be written as~\cite{Miettinen:1978jb,Kowalski:2006hc}
\begin{equation}
\label{eq:coherent}
\frac{\der \sigma^{\gamma^* N \to V N}}{\der t} = \frac{1}{16\pi} \left| \langle \A^{\gamma^* N \to V N}(\xpom,Q^2,\boldsymbol{\Delta}) \rangle \right|^2,
\end{equation}
where the average $\langle \rangle$ refers to the average over all possible target configurations, and $N$ can be a proton or a nucleus. 

The scattering amplitude $\A^{\gamma^* N \to V N}(\xpom,Q^2,\boldsymbol{\Delta})$ can be written as a Fourier transform from coordinate space to momentum space~\cite{Kowalski:2006hc},
\begin{multline}
\label{eq:diff_amp}
 \A^{\gamma^* N \to V N}_{T,L}(\xpom,Q^2, \boldsymbol{\Delta}) = i\int \der^2 \rt \int \der^2 \bt \int \frac{\der z}{4\pi}  \\ 
 \times (\Psi^*\Psi_V)_{T,L}(Q^2, \rt,z) \\
 \times e^{-i[\bt - (1-z)\rt]\cdot \boldsymbol{\Delta}}  2N(\bt,\rt,\xpom).
\end{multline}
Here, the two-dimensional vector $\Deltat$ is the transverse momentum transfer to the target, with $|\Deltat| \approx \sqrt{-t}$. The transverse momentum transfer $\Deltat$ is actually the Fourier conjugate to $\bt - (1-z)\rt$ because of the contribution from the non-forward vector meson wave function~\cite{Bartels:2003yj,Kowalski:2006hc}. The impact parameter $\bt$ points to the center of the dipole from the center of the target. The virtual photon to quark-antiquark dipole splitting is  described by the  virtual photon wave function $\Psi$, and the formation of a vector meson is encoded in the vector meson wave function $\Psi_V$. In this work we use the Boosted Gaussian wave function parametrization from Ref.~\cite{Kowalski:2006hc} (see also Ref.~\cite{Chen:2016dlk}).  Note, that the limited knowledge of the vector meson wave function leads to significant uncertainties in the absolute normalization of the cross section (up to $\sim 50\%$), see e.g. Ref.~\cite{Lappi:2013am}. The interaction of a dipole with transverse size $\rt$ and impact parameter $\bt$ with the target (proton or nucleus) is described in terms of the dipole amplitude $N(\rt,\bt,\xpom)$. The target structure is probed at Bjorken-$x$
\begin{equation}
\xpom = \frac{Q^2+M_{V}^2-t}{Q^2+W^2-m_N^2},
\end{equation}
which can be interpreted as the longitudinal momentum fraction of the proton carried by the color-neutral ``pomeron'' exchanged with the diffractively produced vector meson ($J/\Psi$ in this work). Here $W$ is the center-of-mass energy in the photon-nucleon scattering and $m_N$ the nucleon mass.

If we calculate the total cross section for diffractive vector meson production and subtract the coherent contribution, we obtain the incoherent diffractive cross section. Following Ref.~\cite{Caldwell:2009ke} (see also \cite{Miettinen:1978jb,Frankfurt:1993qi,Frankfurt:2008vi,Lappi:2010dd}) the incoherent cross section becomes
\begin{align}\label{eq:incoherent}
\frac{\der \sigma^{\gamma^* N \to V N^*}}{\der t} = \frac{1}{16\pi} &\left( \left\langle \left| \A^{\gamma^* N \to V N}(\xpom,Q^2,\boldsymbol{\Delta})  \right|^2 \right\rangle \right. \notag\\ 
& ~~~ - \left. \left| \langle \A^{\gamma^* N \to V N}(\xpom,Q^2,\boldsymbol{\Delta}) \rangle \right|^2 \right)\,.
\end{align} 
As the incoherent cross section is a variance, it measures the amount of fluctuations in the diffractive scattering amplitude. Additionally, the coherent cross section depends on the average scattering amplitude and consequently on the average target structure. These two cross sections then make it possible to constrain the event-by-event fluctuating structure of the hadron, as shown in case of protons in Ref.~\cite{Mantysaari:2016ykx} and with heavy nuclei in Ref.~\cite{Mantysaari:2017dwh}.

There are two phenomenological corrections to the results presented above (see Ref.~\cite{Kowalski:2006hc}). First, in the dilute limit where two gluons are exchanged, one actually probes a two-gluon distribution of the target. A dominant contribution to the cross section then originates from the configuration where one of the gluons is very soft, and in this limit one can relate the result to the standard collinear factorization gluon distribution function by introducing the skewedness correction~\cite{Shuvaev:1999ce}. However, applicability of this correction in the saturation region is not clear.
The second correction originates from the fact that usually one assumes the dipole scattering amplitude to be purely real. Both of these contributions mostly affect the overall normalization of the cross section (the $t$ and energy dependencies are weak, see e.g.~\cite{Mantysaari:2017dwh}). As the overall normalization has a large uncertainty  originating from the poorly constrained vector meson wave function, we only add these corrections approximatively. The skewedness correction is estimated as a $40\%$ increase to the cross section. A similar $10\%$ real part correction  is applied to the results obtained using the IPsat parametrization where the dipole amplitude is purely real. 

\section{Dipole-target scattering}
\label{sec:dipoletarget}
We consider two different descriptions for the dipole-target interaction that allow us to obtain the dipole amplitude $N(\rt,\bt,\xpom)$. These are the IPsat parametrization, in which the geometry does not evolve in $\xbj$ and the Bjorken-$x$ dependence of the saturation scale $\qs^2$ (or density) is parametrized by fitting to HERA data. This simple parametrization is compared with an explicit Color Glass Condensate framework calculation in which the energy evolution of the Wilson lines (and consequently the target geometry) is obtained by solving perturbative evolution equations. Summaries of these two approaches are presented below.

\subsection{IPsat model}
\label{sec:ipsat}
In the IPsat parametrization the saturation scale $\qs^2$ depends on the impact parameter, and the dipole-proton scattering amplitude is written as
\begin{equation}
N^p(\rt,\bt,x) = 1-\exp \left[ -\rt^2 F(x,\rt) T_p(\bt)\right],
\end{equation}
where the transverse density profile function is assumed to be Gaussian: $T_p(\bt)=\frac{1}{2\pi B_p} e^{-\bt^2/(2B_p)}$. The function $F$ contains the DGLAP evolved gluon distribution $xg$:
\begin{equation} 
	F(x,\rt^2) = \frac{\pi^2}{2\nc} \as\left( \mu_0^2 + \frac{C}{r^2} \right)  xg \left(x,  \mu_0^2 + \frac{C}{r^2} \right).
\end{equation}
The free parameters of the model ($\mu_0^2$, $C$, $B_p$ and the initial condition for the DGLAP evolution of $xg$) are fixed by fitting the HERA data in Ref.~\cite{Rezaeian:2012ji} (see also Ref.~\cite{Mantysaari:2018nng}). 

In Refs.~\cite{Mantysaari:2016ykx,Mantysaari:2016jaz}, this parametrization was generalized to the case where the proton has a fluctuating substructure consisting of $N_q$ ``hot spots'' by replacing
\begin{equation}
	T_p(\bt) \to \frac{1}{N_q} \sum_{i=1}^{N_q} T_q(\bt-\bti)  
\end{equation}
with
\begin{equation}
    T_q(\bt) = \frac{1}{2\pi B_q} e^{-\bt^2/(2B_q)}.
\end{equation}
Here $B_q$ is the Gaussian width of each hot spot. The locations of the hot spots are sampled from a Gaussian distribution with width $B_{qc}$. These parameters are constrained as in Ref.~\cite{Mantysaari:2016jaz} by requiring a simultaneous description of the coherent and incoherent $J/\Psi$ photoproduction data from HERA~\cite{Alexa:2013xxa} at $W=75\gev$. Unlike in Ref.~\cite{Mantysaari:2016jaz}, the sampled hot spot locations are shifted to keep the center of mass of the nucleon at the origin. This keeps the deuteron and helium sizes unchanged when the fluctuations are included, but effectively makes the nucleons smaller and consequently the used parameters deviate slightly from the ones used in Ref.~\cite{Mantysaari:2016jaz}. In case of the IPsat parametrization,  we use parameters $B_{qc}=4.5\gev^{-2}$ and $B_q=1.0\gev^{-2}$.

The dipole-proton scattering amplitude $N$ discussed above can be generalized to the dipole-nucleus case as
\begin{equation}
\label{eq:dipole_nuke}
N^A(\rt,\bt,x ) = 1 - \prod_{i=1}^A \left[1 - N^p(\rt, \bt-\bt_i, \xpom) \right],
\end{equation}
where $\bt_i$ are the transverse positions of the nucleons, sampled from the nucleus wave function discussed in Sec.~\ref{sec:nucstruct}. 
Within our framework, this procedure is equivalent to summing the density profiles of the nucleons to obtain that of the nucleus.

Several pieces of data prefer the presence of additional fluctuations of the normalization of $Q_s^2$ \cite{Schenke:2013dpa,McLerran:2015qxa,Mantysaari:2016ykx,Mantysaari:2016jaz}.
We thus allow the (squared) saturation scale $Q_s^2$ of the individual hot spots to fluctuate around its expectation value $\langle Q_s^2 \rangle$ following a log-normal distribution
\begin{equation}
	P(\ln (Q_s^2 / \langle Q_s^2\rangle) ) = \frac{1}{\sqrt{2\pi}\sigma} \exp\left[ -\frac{\ln^2 (Q_s^2/ \langle Q_s^2 \rangle)}{2\sigma^2} \right]\,.
\end{equation}
Here we use $\sigma=0.65$ adjusted to get a better description of the small-$|t|$ part of the incoherent $J/\Psi$ photoproduction off protons as measured by HERA  (in Refs.~\cite{Mantysaari:2016ykx,Mantysaari:2016jaz} with the IPsat parametrization where we did not shift center-of-mass to origin, we used comparable value $\sigma=0.5$
determined in Ref.~\cite{McLerran:2015qxa} based on observed multiplicity fluctuations in proton-proton collisions).
The sampled saturation scales are then re-scaled to keep the average $\langle Q_s^2 \rangle$ unchanged, as the expectation value of the log-normal distribution is not $\langle Q_s^2 \rangle$, see the discussion in Ref.~\cite{Mantysaari:2016jaz}. 

\subsection{Color Glass Condensate}\label{sec:cgc}

The eikonal propagation of a quark through the color field of the target at transverse coordinate $\xt$ is determined by the Wilson lines $V(\xt)$, which describe the color rotation of the quark state.
The Wilson lines at each point in the transverse plane are obtained from the Color Glass Condensate effective theory calculation~\cite{Krasnitz:1998ns,Lappi:2003bi,Schenke:2012wb,Schenke:2013dpa}. The details of the computation, summarized below, are exactly the same as in Refs.~\cite{Mantysaari:2016ykx,Mantysaari:2016jaz}.

The target's color charge densities $\rho^a(\xt)$ in the transverse plane are assumed to be random, and sampled from the McLerran-Venugopalan (MV) model\,\cite{McLerran:1993ni}, where $\langle \rho^a(x^-, \xt)\rangle = 0$, and
\begin{equation}\label{eq:MVrho}
\langle \rho^a(x^-, \xt) \rho^b(y^-,\yt) \rangle =  \delta^{ab} \delta^{(2)}(\xt-\yt) \delta(x^- - y^-) g^2 \mu^2\,,
\end{equation}
where $a$ and $b$ are color indices.
The color charge density  $g\mu$ is proportional to the saturation scale $Q_s(\xpom, \xt, \bt)$ as $Q_s = c g^2 \mu$, with $c= 0.7$ when nucleon shape fluctuations are included and and $c=0.65 g^2 \mu$ without, as in Ref.~\cite{Mantysaari:2016jaz} (see also Ref.~\cite{Lappi:2007ku}).
The saturation scale $Q_s^2$, which is a proxy for the nucleon density, is obtained from the IPsat parametrization of the dipole amplitude, presented previously. It is defined via the relation
\begin{equation}
	N\left(\rt^2 = 2/Q_s^2, \bt, x\right) = 1 - e^{-1/2}.
\end{equation}

After the saturation scale at every point in the transverse plane, $\qs^2(\bt)$, is obtained, we can sample color charges according to \eqref{eq:MVrho} and solve the Yang-Mills equations to determine the Wilson lines at every transverse position~\cite{Schenke:2012wb}:
 \begin{equation}\label{eq:wilson}
  V (\xt) = P \exp\left({-ig\int \der x^{-} \frac{\rho(x^-,\xt)}{\boldsymbol{\nabla}^2+\tilde m^2} }\right)\,.
\end{equation}
Here $P$ indicates path ordering. In order to suppress long-distance Coulomb tails, an effective mass regulator  $\tilde m$ is introduced. In general one expects $\tilde m \sim \lqcd$, and here we use $\tilde m=0.4\gev$ as constrained in Ref.~\cite{Mantysaari:2016jaz} by comparing with the HERA $J/\Psi$ photoproduction data.

All nucleon shape and density fluctuations are included via $Q_s^2(\bt)$ calculated from the IPsat model, where geometry fluctuations are included as discussed in Sec.~\ref{sec:ipsat}. Here we use the parameters $B_{qc}=4.0\gev^{-2}$ and $B_q=0.3\gev^{-2}$, and $\sigma=0.5$ to determine the magnitude of the $Q_s^2$ fluctuations, as in Ref.~\cite{Mantysaari:2016ykx}.

After the Wilson lines at the initial $x_0=0.01$ are sampled as discussed above, the evolution towards smaller $x$ is obtained by solving the JIMWLK renormalization group equation~\cite{JalilianMarian:1996xn,JalilianMarian:1997jx,JalilianMarian:1997gr,Iancu:2001md}. Here we use exactly the same setup as in Ref.~\cite{Mantysaari:2018zdd}, which we summarize below.

The JIMWLK equation describes the rapidity evolution of a Wilson line and can be written as a stochastic Langevin equation ~\cite{Blaizot:2002np}
\begin{equation}\label{eq:langevin1}
\frac{\der}{\der y} V(\xt) = V(\xt) (i t^a) \left[
\int \der^2 \zt\,
\varepsilon_{\xt,\zt}^{ab,i} \; \xi_\zt(y)^b_i  + \sigma_\xt^a 
\right]\,,
\end{equation}
where $t^a$ is an SU(3) generator in the fundamental representation.
The evolution \eqref{eq:langevin1} can be seen as a random walk in the color space, where the random noise $\xi$ is a local Gaussian variable with variance
\begin{equation}
\label{eq:noice}
\langle \xi_{\xt,i}^a(y) \xi_{\yt,j}^b(y')\rangle = \delta^{ab} \delta^{ij} \delta^{(2)}_{\xt\yt} \delta(y-y').
\end{equation}
The coefficient of the noise is
\begin{equation}
\label{eq:noicecoef}
 \varepsilon_{\xt,\zt}^{ab,i} = \left(\frac{\as}{\pi}\right)^{1/2}\;
K_{\xt-\zt}^i
\left[1-V_\xt^\dag  V_\zt\right]^{ab},
\end{equation}
and the kernel reads
\begin{equation}
K_\xt^i = \frac{x^i}{\xt^2}.
\end{equation}
The deterministic drift term $\sigma_\xt^a$ can be eliminated from the equation following Ref.~\cite{Lappi:2012vw}, which avoids the requirement to evaluate Wilson lines in the adjoint representation (as appearing in \eqref{eq:noicecoef}) and makes the numerical solution more efficient.

The JIMWLK kernel $K_\xt^i$ has a powerlike structure, and consequently the evolution generates long-distance Coulomb tails (similar to those regulated at the initial condition by the mass parameter $\tilde m$ in Eq.~\eqref{eq:jimwlk_m}). 
This would result in an exponentially growing cross section with rapidity \cite{GolecBiernat:2003ym,Schlichting:2014ipa}, and violate the Froissart bound \cite{Kovner:2001bh}. To avoid this we follow Ref.~\cite{Schlichting:2014ipa} and introduce effective confinement scale effects by using a modified kernel
\begin{equation}
\label{eq:jimwlk_m}
K_\xt^i \to m|\xt| K_1(m|\xt|) \frac{x^i}{\xt^2}.
\end{equation}
Here the modified Bessel function $K_1$ suppresses contributions at distance scale $\gtrsim 1/m$. In this work we use $m=0.2\gev$ as constrained in~\cite{Mantysaari:2018zdd} to be compatible with the HERA structure function data when the evolution is performed at fixed coupling $\as=0.21$. In Ref.~\cite{Mantysaari:2018zdd} we showed that the diffractive cross sections are sensitive to the infrared regulator $m$ only at small $t$ (if the strong coupling constant is adjusted to keep the evolution speed compatible with the structure function data).

Calculations are performed on a 2-dimensional lattice with transverse spacing $a=0.01\,{\rm fm}$ and $L=13$ fm in case of deuterons and $L=10$ fm with helium (note that larger lattices are needed with deuterons to accurately describe the tail of the distribution shown in Fig.~\ref{fig:deuteron_size}). We have checked that smaller lattice spacings do not alter the results. For more details on the JIMWLK evolution and its implementation on a lattice, the reader is referred to Ref.~\cite{Mantysaari:2018zdd}.

\section{Energy evolution of light nuclei}
\label{sec:evolution}
\begin{figure}[tb]
\centering
		\includegraphics[width=0.48\textwidth]{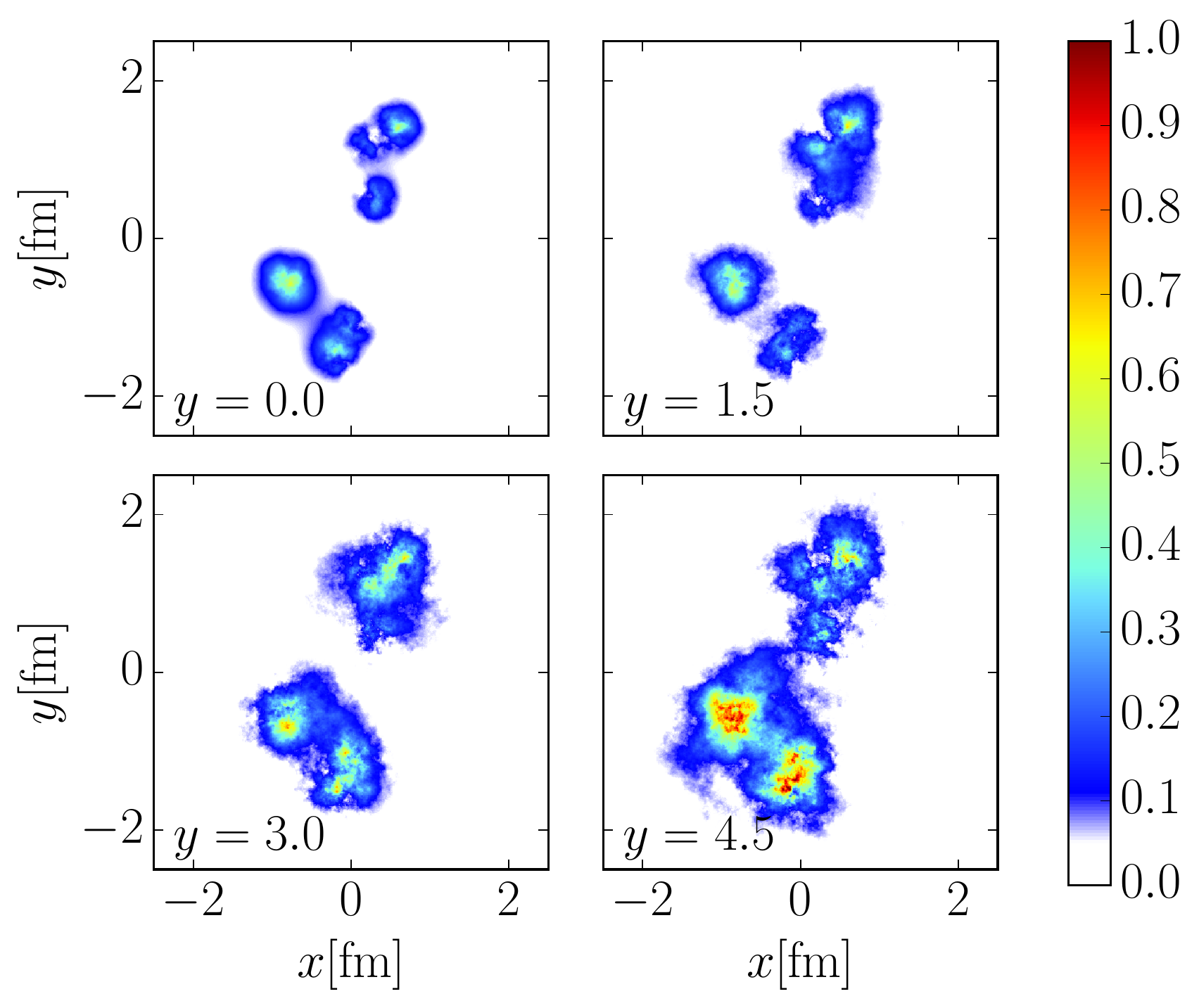} 
				\caption{Illustration of the deuteron density profile and its evolution in the case where nucleon shape fluctuations are included. The density is represented by $1 - \mathrm{Re} \tr V / N_\text{c}$. }
		\label{fig:deuteron_density_evolution}
\end{figure}

\begin{figure}[tb]
\centering
		\includegraphics[width=0.48\textwidth]{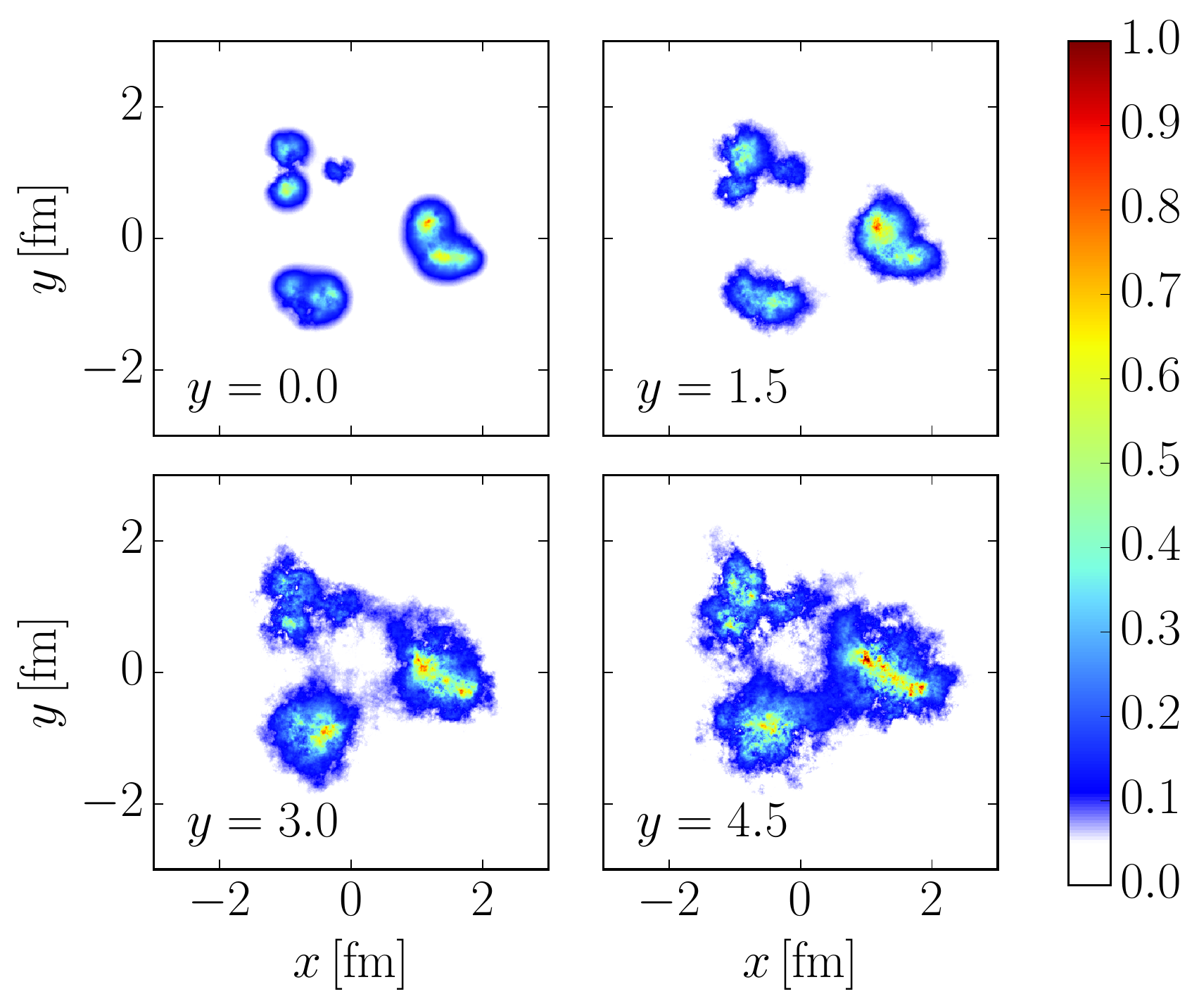} 
				\caption{Illustration of the helium-3 density evolution with nucleon shape fluctuations. The density is represented by $1 - \mathrm{Re} \tr V / N_\text{c}$.}
		\label{fig:helium_density}
\end{figure}

In this section we consider the CGC framework of Sec.~\ref{sec:cgc} and illustrate the evolution of individual deuteron and $^3$He configurations. Results for the deuteron including proton and neutron structure fluctuations and evolution over a few units of rapidity (note that $x = x_0 e^{-y}$ with $x_0=0.01$ in this work) are shown in Fig.~\ref{fig:deuteron_density_evolution}. The deuteron density is characterized by the trace of the Wilson line $1 - \mathrm{Re} \tr V / N_\text{c}$. The evolution first washes out the proton substructure (as already noted in Ref.~\cite{Schlichting:2014ipa}), and eventually the nucleons grow enough to create one large region of gluon matter. However, with a typical proton-neutron transverse separation $\sim 2$ fm one needs to go to very low $x$ in order to see the two nucleons merging.
Similarly, in case of ${}^3$He, evolution for one example configuration (with nucleon substructure) is shown in Fig.~\ref{fig:helium_density} where very similar effects can be seen.

To quantify how the presence of two nucleons affects the deuteron evolution in comparison to the single nucleon case, we study the deuteron density evolution as a function of rapidity $y = \ln \frac{x_0}{x}$. 
In this case we fix the deuteron transverse size to a typical value $d_\text{pn}=1.5\fm$ and quantify the density by calculating the average dipole-deuteron scattering amplitude with a fixed size dipole with $|\rt|=0.2\fm$. Note that this is a typical scale for the dipoles contributing to the $J/\Psi$ production~\cite{Kowalski:2006hc}. 

For comparison, we show the case where we consider the deuteron to consist of two independently evolved nucleons. In that case, the scattering amplitude $N_{pn}$ for the proton neutron system at point $b$ (where $b=0$ is the center of the deuteron) is obtained following Eq.~\eqref{eq:dipole_nuke}:
\begin{multline}
N_{pn}(r,b) = N(r, |b-d_{pn}/2|) + N(r, |b+d_{pn}/2|) \\- N(r,|b-d_{pn}/2|)N(r,|b+d_{pn}/2|),
\end{multline}
where $N=N^p$ is the dipole-nucleon scattering amplitude.

\begin{figure}[tb]
\centering
		\includegraphics[width=0.48\textwidth]{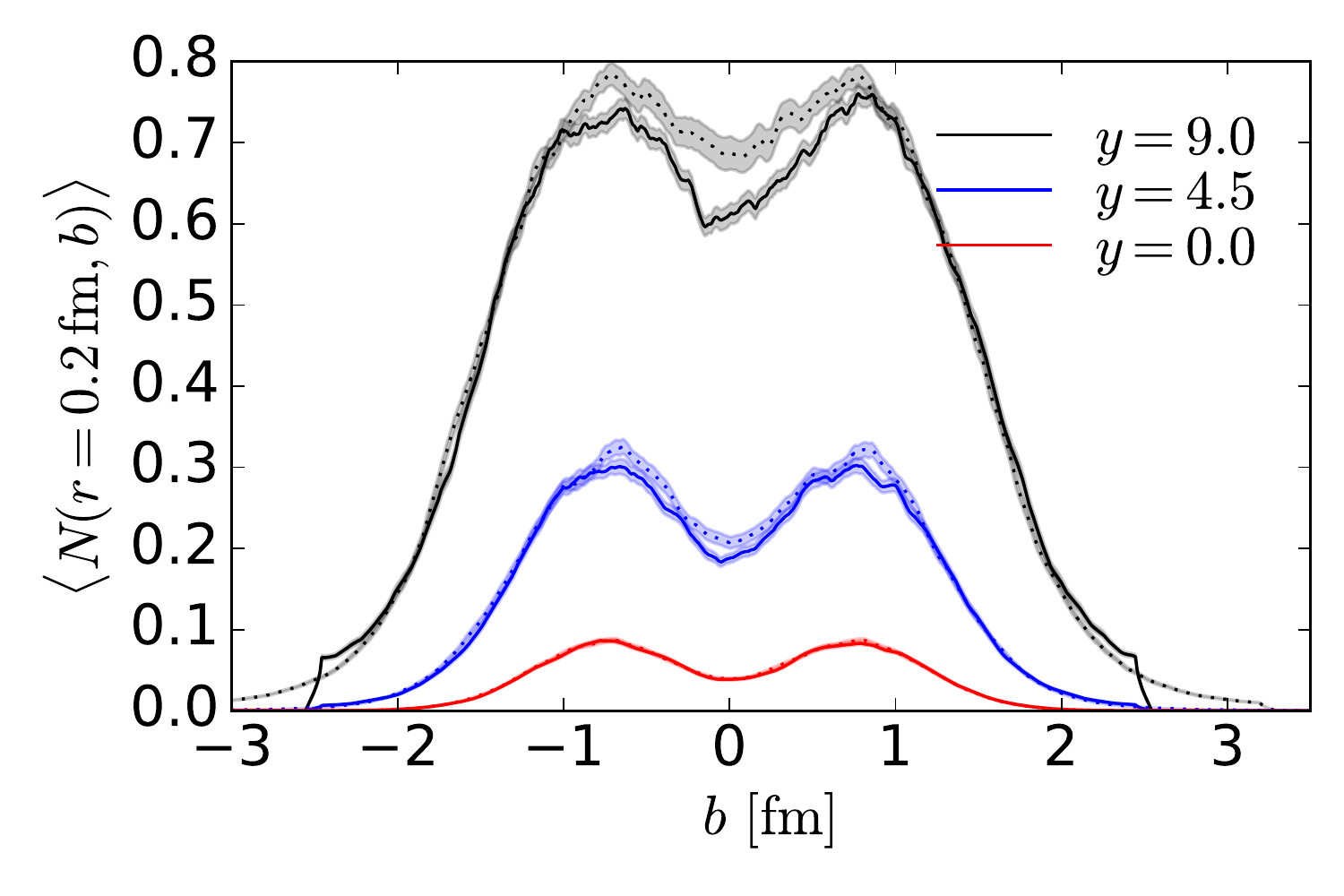}
		\caption{Deuteron density measured by the fixed size dipole as a function of the impact parameter. The proton-neutron separation is $d_\text{pn}=1.5 \fm$. The dashed line shows the result in the case where the proton and neutron are evolved independently. }
		\label{fig:deuteron_dipole_evolution_b}
\end{figure}

The impact parameter dependence of the dipole amplitude is shown in Fig.~\ref{fig:deuteron_dipole_evolution_b}, where the solid lines refer to the deuteron and dotted lines assume independently evolving nucleons.
We find that at the beginning of the evolution the two nucleons are separated enough spatially and the deuteron evolution is very close to the independent nucleon case. 
Only at large rapidities $y\gtrsim 4$ we start to observe small deviations. We note that the rapidity at which this deviation begins depends heavily on the deuteron size $d_\text{pn}$, and for smaller deuterons enhanced non-linear effects in the dense region start to slow down the evolution earlier.

\begin{figure}[tb]
\centering
		\includegraphics[width=0.48\textwidth]{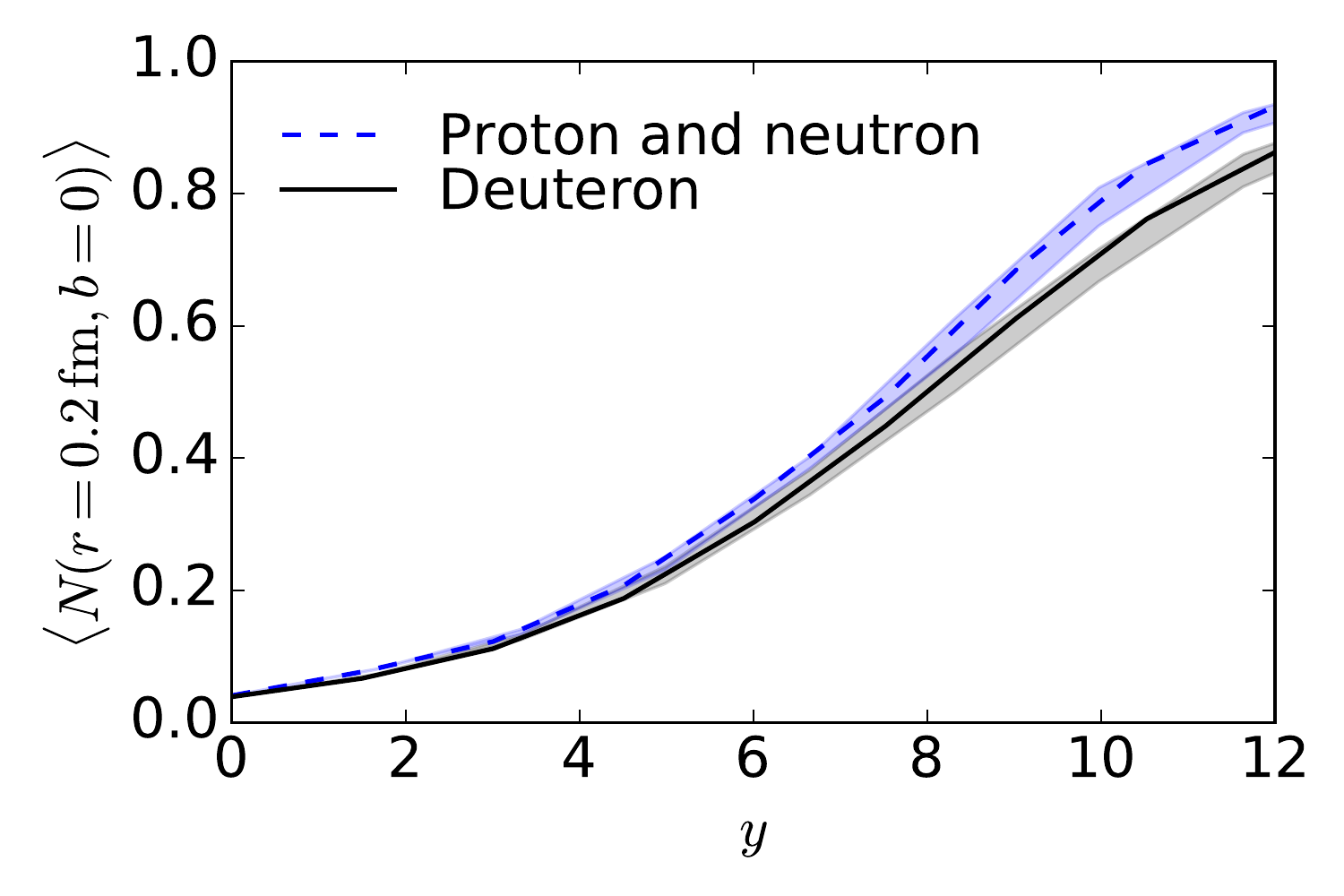} 
				\caption{Evolution of the deuteron density at the origin for a proton-neutron separation of $d_\text{pn}=1.5\,\mathrm{fm}$ (solid), compared to the case where the two nucleons evolve independently (dashed). 
				The density is measured as the average interaction strength with the fixed size dipole.  The lines are our standard choice with infrared regulator $m=0.2\gev$, and the band reflects the variation when $m$ is changed from $0.1\gev$ (lower) 
				to $0.4\gev$ (upper). 
				}
		\label{fig:deuteron_dipole_evolution}
\end{figure}

The rapidity dependence of the dipole amplitude at zero impact parameter is shown in Fig.~\ref{fig:deuteron_dipole_evolution}. The evolution is identical in both cases until the proton and neutron grow so much that they start to overlap significantly at $y\gtrsim 5$, when the non-linear effects start to decrease the evolution speed slightly compared with the independent nucleon case.  

To demonstrate that our analysis is insensitive to the infrared regulator $m$, we vary $m$ in the JIMWLK kernel~\eqref{eq:jimwlk_m} and extract the same dipole amplitude evolution at zero impact parameter. The value of the strong coupling constant for each infrared regulator is constrained in \cite{Mantysaari:2018zdd} from fits to HERA data. 
Besides our standard choice $m=0.2\gev$ and $\as=0.21$, we use $m=0.1\gev$ with $\as=0.19$ and $m=0.3\gev$ with $\as=0.225$. As can be seen from Fig.~\ref{fig:deuteron_dipole_evolution}, the evolution is very similar in all three cases. This was not entirely obvious, because the scenario of smaller $m$ and smaller $\as$ for example, leads to faster evolution of the low momentum modes, which dominate the evolution of the nucleon size. For larger dipoles the sensitivity to infrared regulators should be enhanced, but contributions from large dipoles to vector meson production is suppressed exponentially.

 \section{Predictions for the EIC}
 \label{sec:predictions}
 We consider $J/\Psi$ production in the kinematical range accessible with a future US Electron Ion Collider, where the center-of-mass energies can reach up to $\sqrt{s_{NN}} = 140\sqrt{Z/A}\gev$~\cite{Aschenauer:2017jsk}. This allows the reach of $\xpom$ values down to $\xpom \sim 10^{-4}\dots 10^{-3}$ in $J/\Psi$ photoproduction in electron-deuteron and electron-helium collisions.
 
 \subsection{Short range nucleon-nucleon correlations in the deuteron}
The structure of light nuclei at low energy scales is well known, but so far it has not been possible to probe the distribution of small-$x$ gluons in e.g. the deuteron wave function. To demonstrate that the future Electron-Ion Collider can do detailed imaging of the small-$x$ gluon distributions, we first study vector meson production off deuterons using two different realistic wave functions to describe the proton-neutron separation in the deuteron, assuming that small-$x$ gluons are distributed around the nucleons. 
 
The difference between the used Hulthen and Argonne wave functions is that the short range nucleon-nucleon correlations that suppress configurations where proton and neutron are close to each other ($d_\mathrm{pn} \lesssim 0.5\fm$) are included in the Argonne wave function, see Fig.~\ref{fig:deuteron_size}. 
For simplicity the nucleon shape fluctuations are not included in this analysis.
 
\begin{figure}[tb]
\centering
		\includegraphics[width=0.48\textwidth]{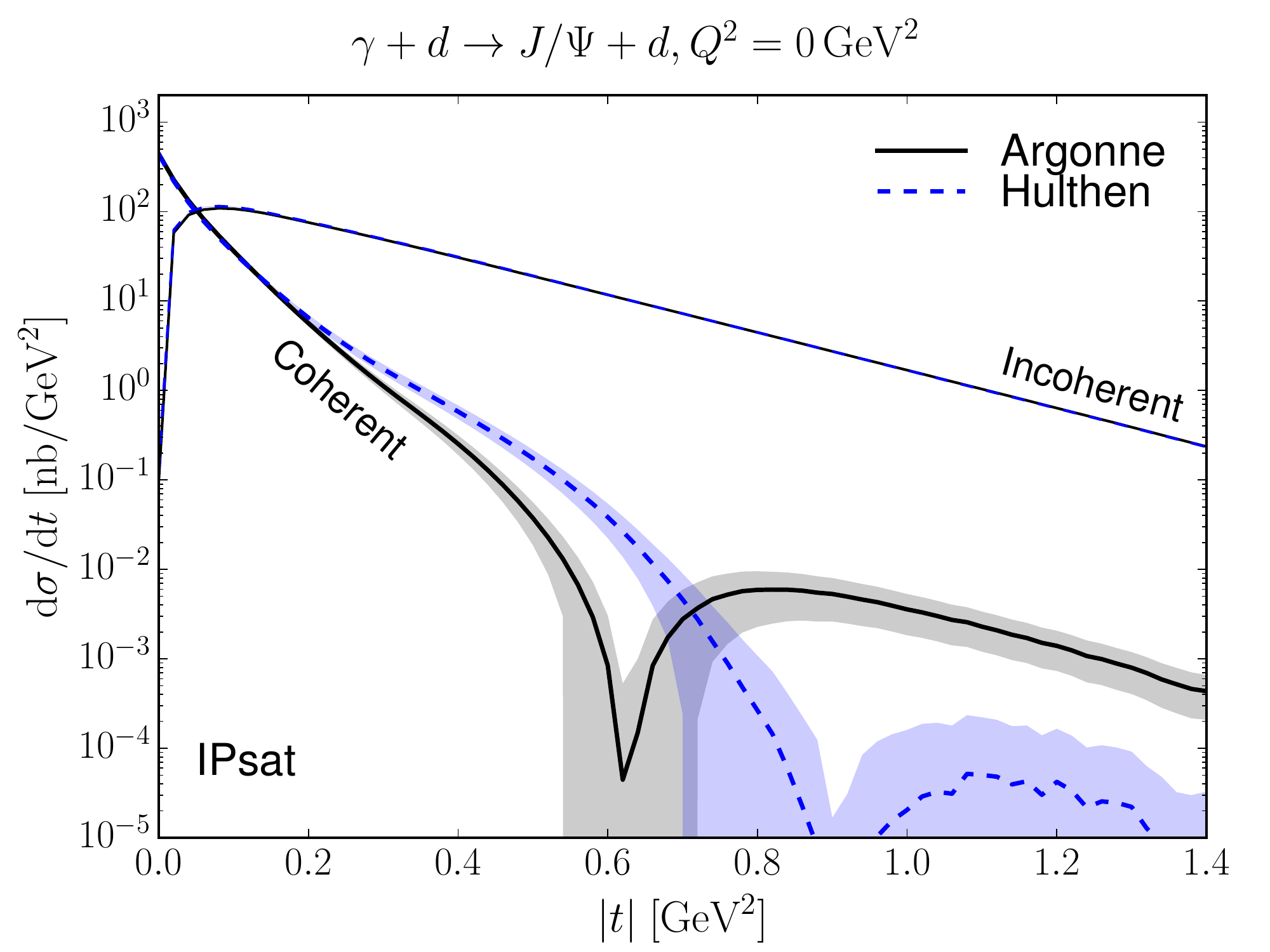} 
				\caption{Coherent and incoherent $J/\Psi$ production cross sections at $\xpom=0.01$  as a function of $t$, with two different wave functions to describe the deuteron structure. Nucleon shape fluctuations are not included.}
		\label{fig:deuteron_hulthen_vmc}
\end{figure}

 The obtained $J/\Psi$ photoproduction cross sections for coherent and incoherent production at fixed $\xpom=0.01$ are shown in Fig.~\ref{fig:deuteron_hulthen_vmc}. Here, we use the IPsat parametrization to describe the dipole-deuteron interaction. The effect of having different average density profiles for the gluon distribution results in coherent cross sections that deviate at $|t|\gtrsim 0.3\gev^2$. At smaller $|t|$, where one is sensitive to the structure at long distance scales, the spectra are identical. 
 
 Generally, the position of the first diffractive minimum corresponds to the size of the target $R$ as $t_\text{dip} \sim 1/R^2$. Here, the root mean square separations for the proton and neutron are similar in both wave functions, with the Hulthen wave function resulting in deuterons that are slightly \emph{larger} by $\sim 3\%$. So the difference in the observed spectra must be due to the different average shape (the $t$ spectra is the Fourier transform of the impact parameter profile, and the density profiles differ as seen in Fig.~\ref{fig:deuteron_size}), not different average size. If the deuteron size is characterized by measuring the slope $B_D$ of the coherent cross section close to $t=0$ (parametrizing the spectra as $\sim e^{-B_D |t|}$), we find that the spectrum is slightly steeper when the Hulthen wave function is used $(B_D \approx 28\gev^{-2}$ with Argonne potential wave function and $B_D\approx 29\gev^{-2}$ with Hulthen), consistent with the slightly larger RMS size.
 
 The incoherent cross sections are basically identical in the studied $|t|$ range, with the largest difference at $|t|\sim 0.1\gev^2$, where the result with the Argonne potential wave function is below the Hulthen result by $\approx 5\%$. This can be understood, as the short-range correlations effectively reduce overall density fluctuations by rejecting some of the configurations where the nucleons overlap in the transverse plane. However, the effect is numerically small and most likely not observable. 
 
\begin{figure}[tb]
\centering
		\includegraphics[width=0.48\textwidth]{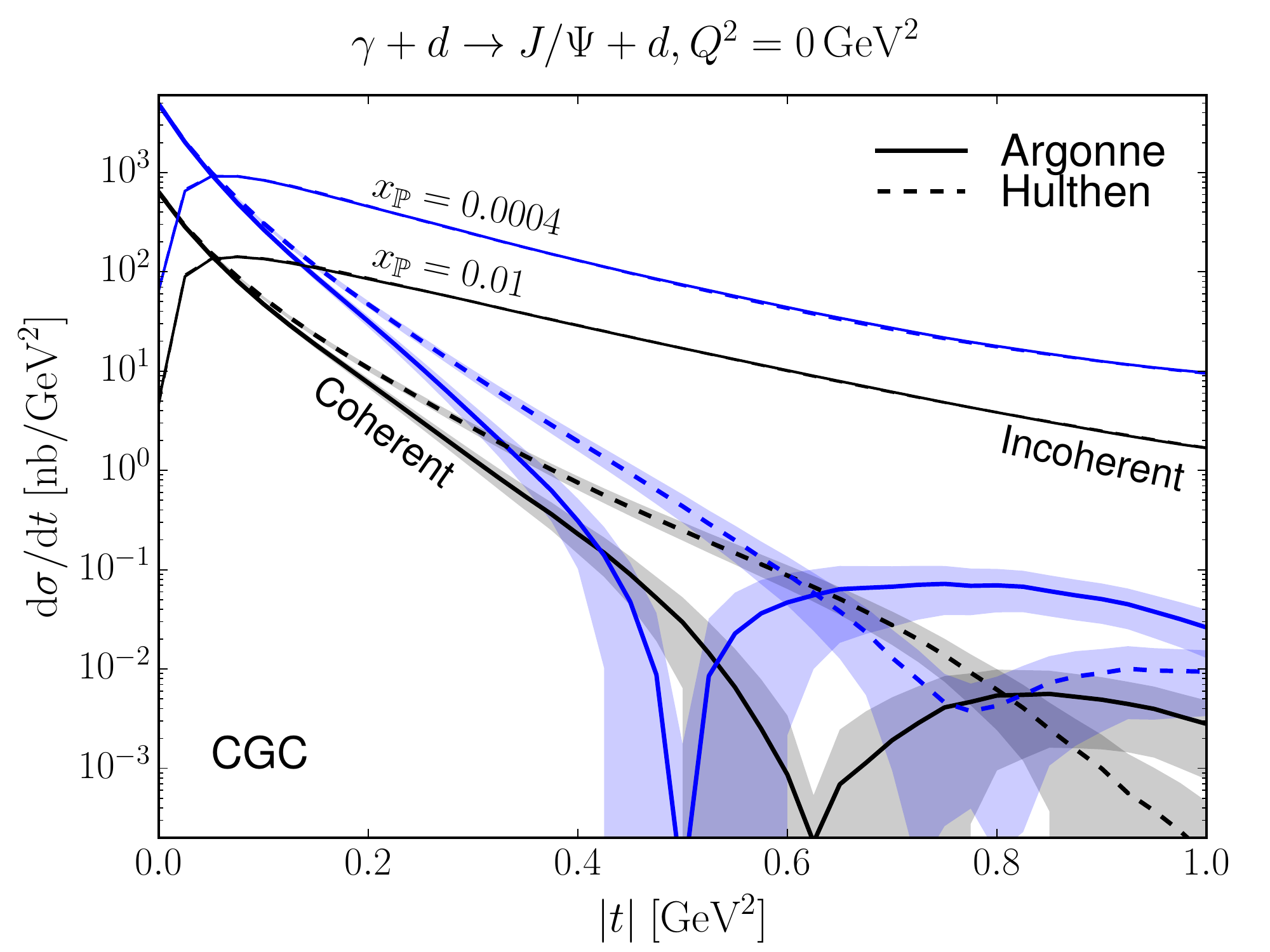} 
				\caption{Coherent and incoherent $J/\Psi$ production cross sections at $\xpom=0.01$ (lower black lines) and $\xpom=0.0004$ (upper blue lines) as a function of $t$, calculated from the CGC famework using two different wave functions for the deuteron at $\xpom=0.01$. Nucleon shape fluctuations are not included in the calculation. }
		\label{fig:deuteron_hulthen_vmc_cgc}
\end{figure}

To study if the difference between the two wave functions for the deuteron is washed out by the small-$x$ evolution, we next show predictions for $J/\Psi$ photoproduction calculated from the CGC framework using both Hulthen and Argonne wave functions to describe the proton-neutron separation at  $\xpom=0.01$. The energy evolution for each configuration is obtained by solving the JIMWLK equation. The resulting spectra at $\xpom=0.01$ and $\xpom=0.0004$ are shown in Fig.~\ref{fig:deuteron_hulthen_vmc_cgc}. Here, we do not include nucleon shape fluctuations for simplicity. Similar to the case of the IPsat parametrization, the coherent cross sections obtained with different wave functions deviate above $|t|\gtrsim 0.2\gev^2$, and the position of the first diffractive minimum is at a smaller $|t|$ when the Argonne wave function is used. This difference is found to remain similar after the JIMWLK evolution down to $\xpom=0.0004$. 
This shows that the small $x$ evolution retains the differences between the deuteron wave functions defined at large $x$ within the $x$ range accessible with a future EIC.

\subsection{Deuteron shape and its small-$x$ evolution}

\begin{figure}[tb]
\centering
		\includegraphics[width=0.48\textwidth]{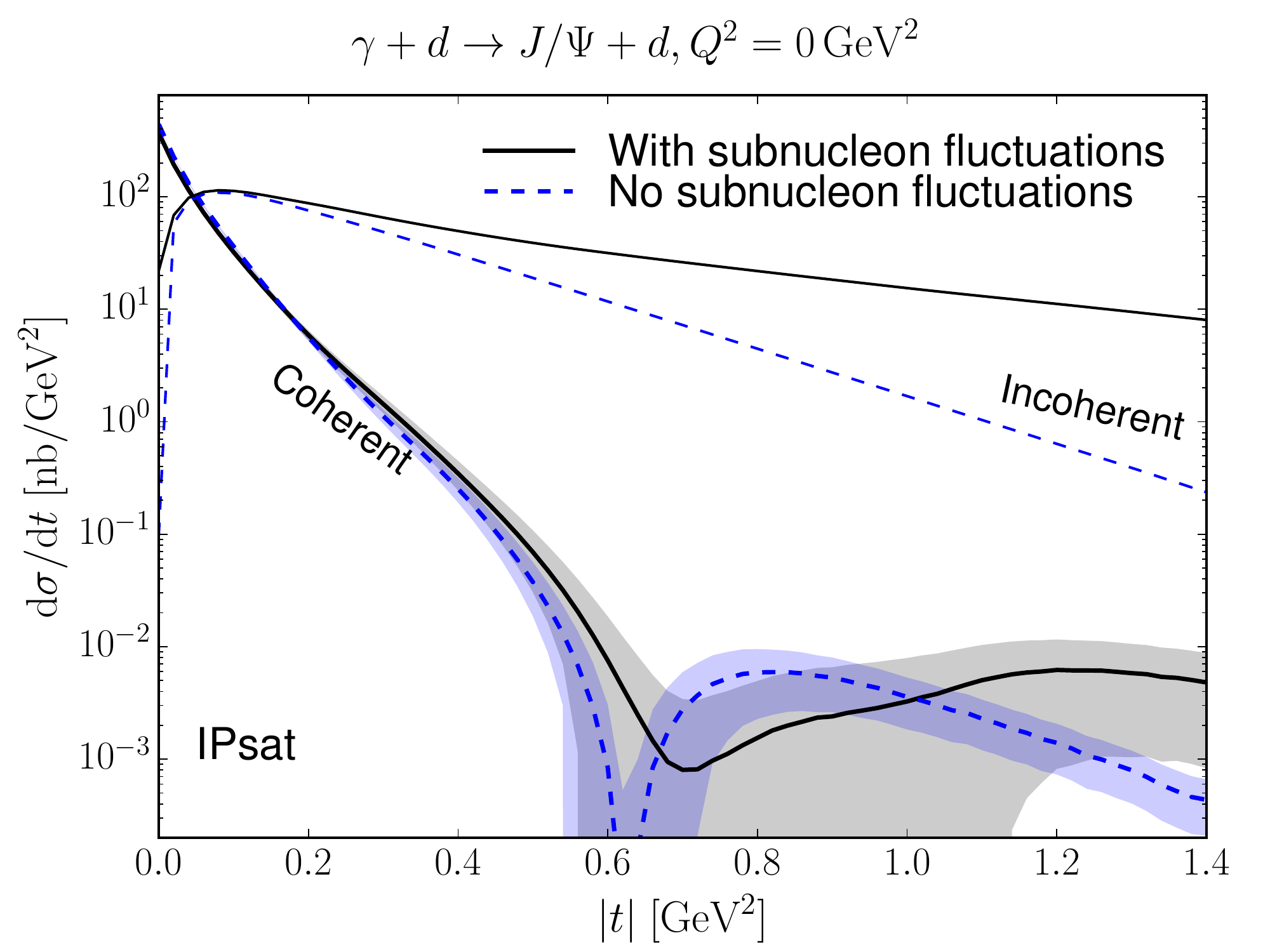} 
				\caption{Coherent and incoherent diffractive $J/\Psi$ photoproduction cross section in electron-deuteron collisions at $\xpom=0.01$, with (solid) and without (dashed) sub-nucleonic fluctuations.    }
		\label{fig:ipsat_w_wo_fluct}
\end{figure}

\begin{figure}[tb]
\centering
		\includegraphics[width=0.48\textwidth]{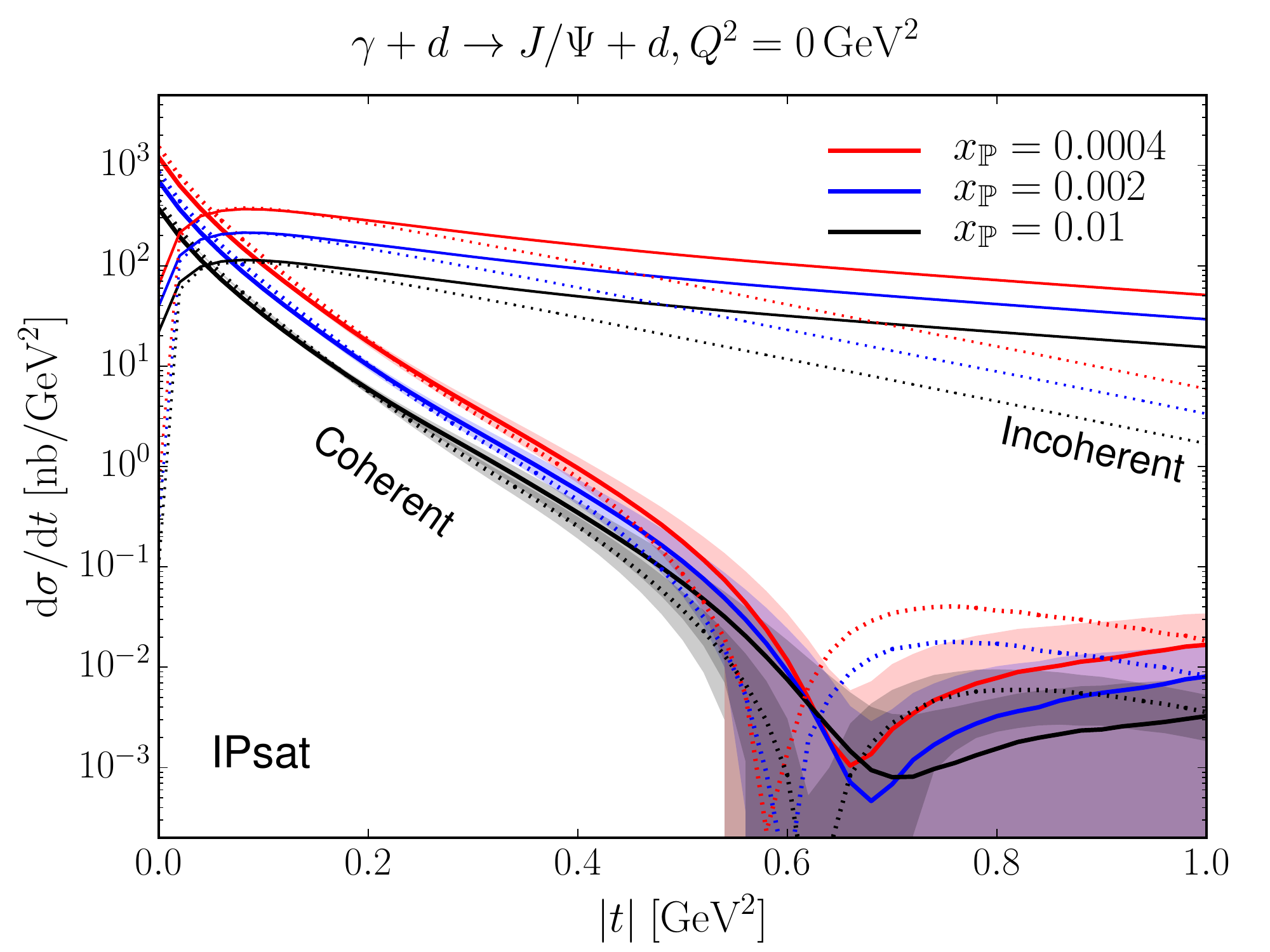} 
				\caption{Coherent and incoherent diffractive $J/\Psi$ photoproduction cross sections in photon-deuteron collisions at different $\xpom$ from the IPsat parametrization. Solid lines include nucleon shape fluctuations and dotted lines do not. For clarity, statistical uncertainty of the calculation is only shown for the case with fluctuating substructure, where errors are much larger.
				}
		\label{fig:ipsat_evolution}
\end{figure}

\begin{figure}[tb]
\centering
		\includegraphics[width=0.48\textwidth]{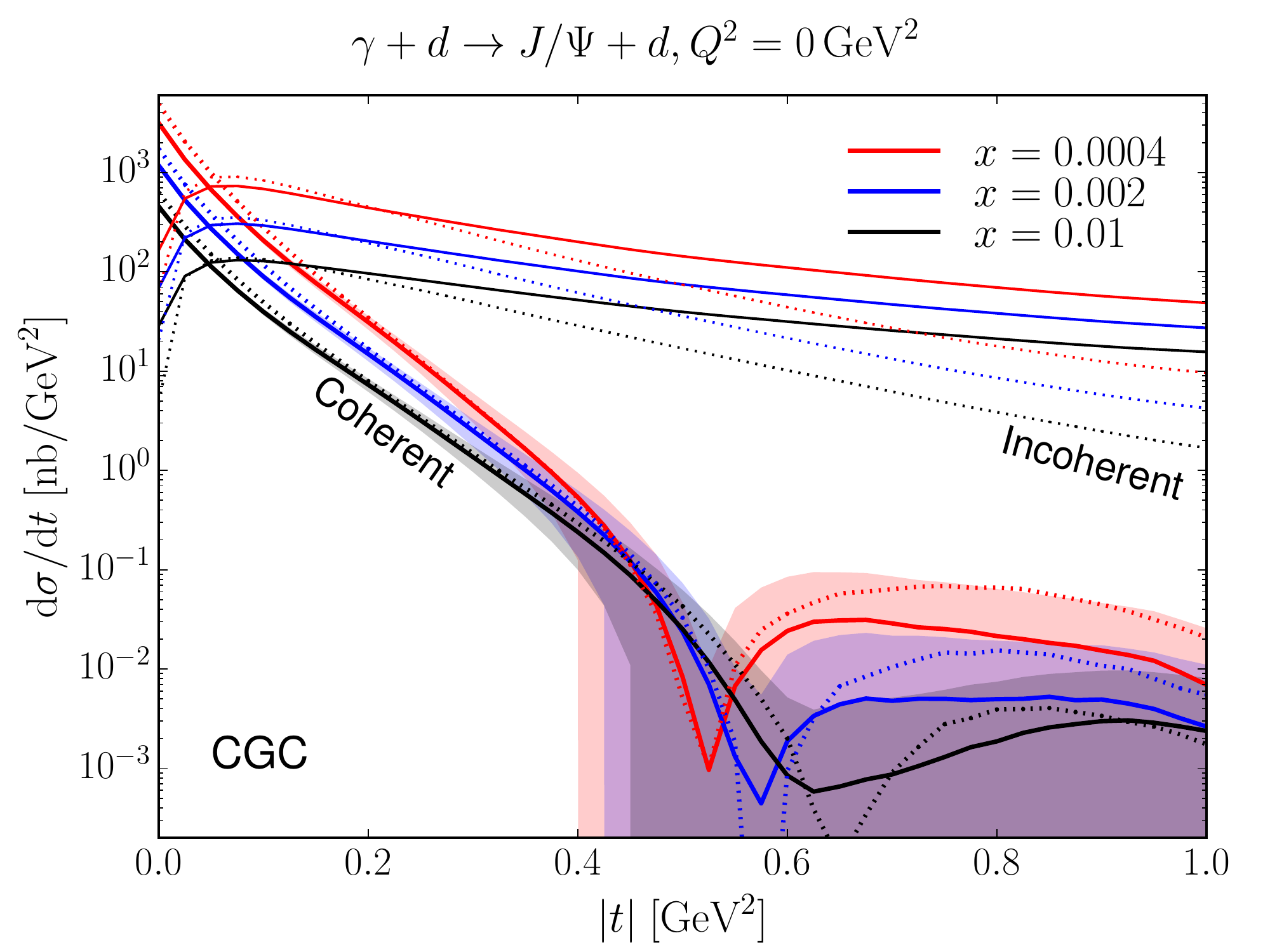} 
				\caption{Coherent and incoherent $J/\Psi$ photoproduction off deuterons calculated from the CGC framework at different $\xpom$. Solid lines include nucleon shape fluctuations and dotted lines do not. For clarity, statistical uncertainty of the calculation is only shown for the case with fluctuating substructure, where errors are much larger.}
		\label{fig:cgc_evolution}
\end{figure}

First we study the effect of nucleon shape fluctuations on  diffractive $J/\Psi$ production off deuterons. In the following we will only use the Argonne potential wave function to describe the proton-neutron separation in the deuteron. Our conclusions would be similar if the Hulthen wave function were used.

Using the IPsat model with and without nucleon shape fluctuations, we calculate the coherent and incoherent $J/\Psi$ photoproduction cross sections with detueron targets at $\xpom=0.01$. The results are shown in Fig.~\ref{fig:ipsat_w_wo_fluct}.
We find that the coherent cross sections are compatible within the numerical accuracy  up to $|t| \gtrsim 1.2\gev^{2}$ with and without subnucleonic fluctuations 
(note that introducing the additional fluctuations approximately leaves the average shape unchanged). 
On the other hand, the incoherent cross sections start to deviate significantly at $|t| \approx 0.25\gev^2$, similar to the case of photon-heavy nucleus scattering analyzed in Ref.~\cite{Mantysaari:2017dwh}. Note that at smaller $|t|$, where we are sensitive to fluctuations at longer length scales (scale of the proton and neutron separation, sampled from the deuteron wave function), there is basically no difference in the incoherent cross sections.

Next, we study the energy dependence of the cross sections. We calculate $J/\Psi$ production at different $\xpom$ values in the EIC energy range, and the results are shown in Fig.~\ref{fig:ipsat_evolution} for the IPsat model, and in Fig.~\ref{fig:cgc_evolution} for the CGC framework, where the JIMWLK evolution equation describes the structure evolution.  The results are shown with nucleon shape fluctuations (solid lines) and without (dotted lines). 

\begin{figure}[tb]
\centering
		\includegraphics[width=0.48\textwidth]{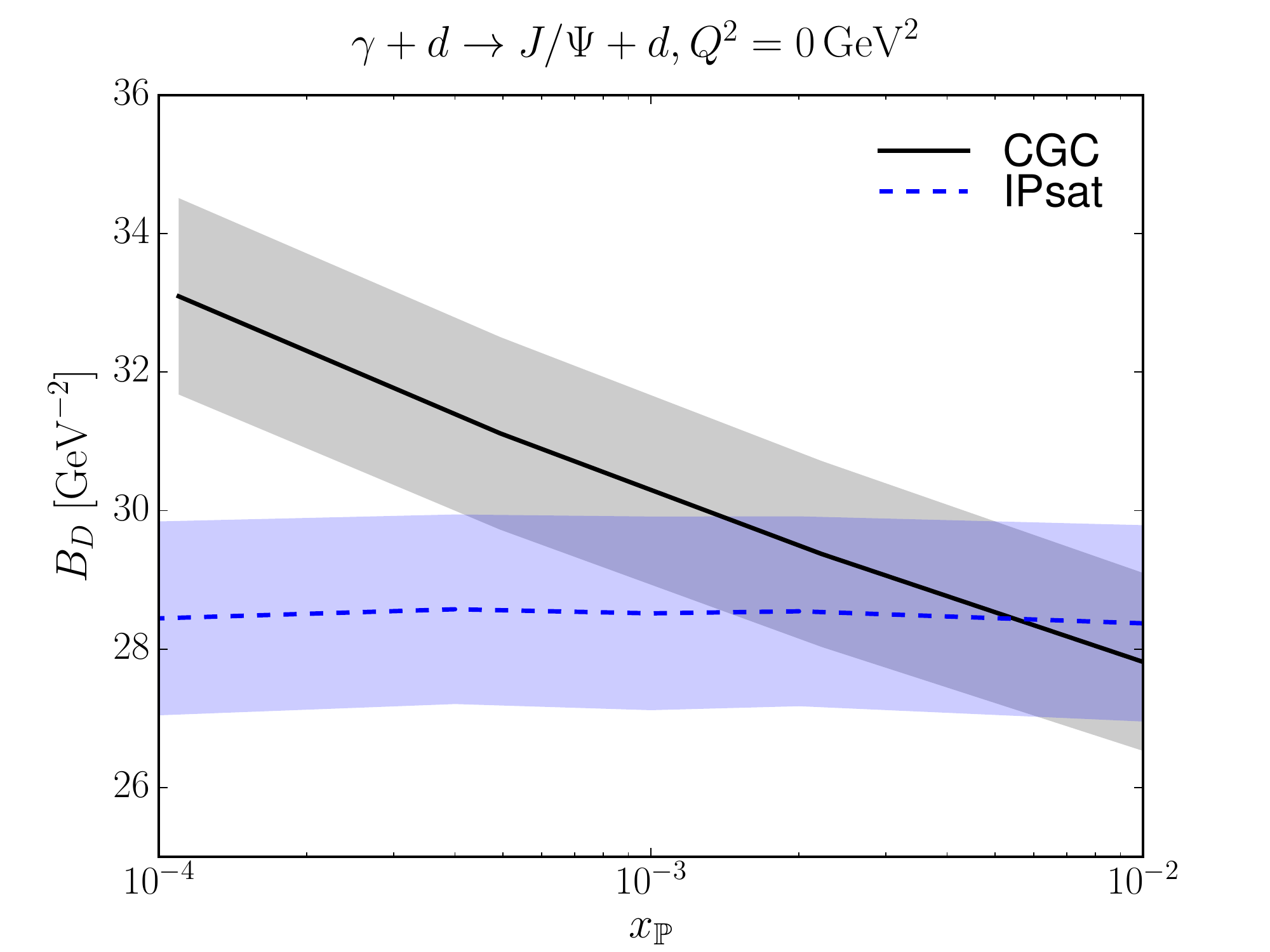} 
				\caption{Slope of the coherent cross section at small $|t|$ extracted from the calculation without nucleon shape fluctuations as a function of $\xpom$. The band shows the statistical uncertainty of the slope extraction.}
		\label{fig:deuteron_slope}
\end{figure}

We find that the position of the first diffractive minimum moves to smaller $|t|$ as a result of the JIMWLK evolution in the Color Glass Condensate calculation, which results in growing deuteron size as illustrated in Fig.~\ref{fig:deuteron_density_evolution}. On the other hand, in case of the IPsat parametrization where there is no geometry evolution the dip location is approximately constant.  Similarly, the coherent cross section at small $|t|$ gets steeper when the JIMWLK evolution is performed and remains constant in the IPsat calculation. This is demonstrated explicitly by parametrizing the coherent cross section as $\der \sigma/\der t \sim e^{-B_D |t|}$ and extracting the $|t|$ slope $B_D$ as a function of $\xpom$, shown in Fig.~\ref{fig:deuteron_slope}.

 \begin{figure}[tb]
\centering
		\includegraphics[width=0.48\textwidth]{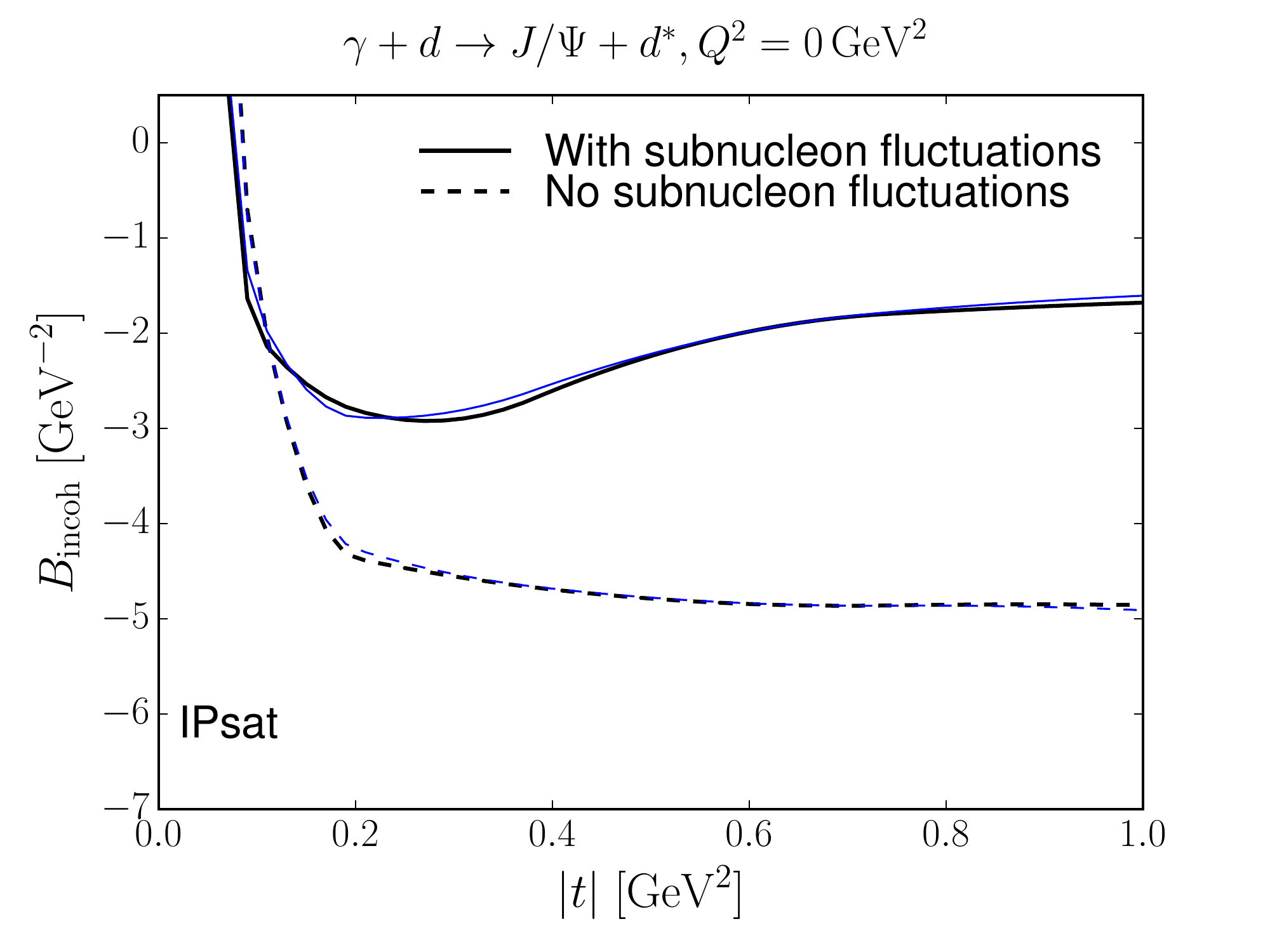} 
				\caption{Slope of the incoherent cross section in $\gamma$+deuteron scattering from IPsat. Black lines are at  $\xpom=0.01$ and blue lines at $\xpom=0.0004$.  }
		\label{fig:deuteron_incoh_slope_ipsat}
\end{figure}

 \begin{figure}[tb]
\centering
		\includegraphics[width=0.48\textwidth]{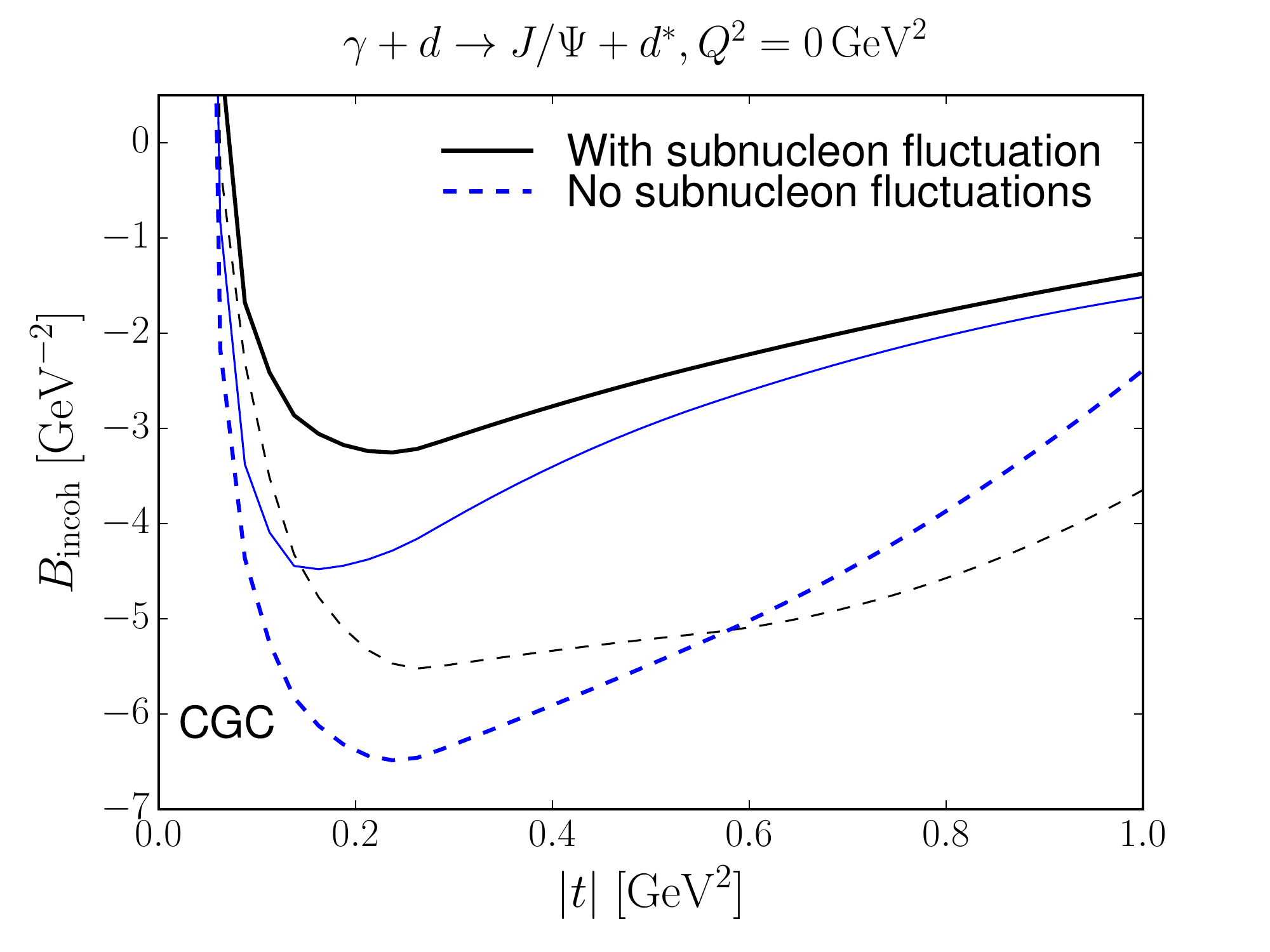} 
				\caption{Slope of the incoherent cross section in $\gamma$+deuteron scattering from the CGC. Black lines are at $\xpom=0.01$ and blue lines at $\xpom=0.0004$.  }
		\label{fig:deuteron_incoh_slope}
\end{figure}

As shown in Ref.~\cite{Lappi:2010dd}, the $|t|$ slope of the incoherent cross section $B_\text{incoh}$ is controlled by the size of the object that is fluctuating. If the cross section is parametrized as $e^{-B_\text{incoh} |t|}$, at moderate $|t|$ the slope is controlled by the size of the nucleon, with $B_\text{incoh} \sim 4 \dots 5 \gev^{-2}$ (note that the nucleon RMS size is set by $B_p=4\gev^{-2}$). At large $|t|$, if the substructure fluctuations are included and the nucleons consist of hot spots, the slope approaches the hot spot size and $B_\text{incoh} \sim 1 \dots 2 \gev^{-2}$ (recall that our hot spot size in the IPsat parametrization is $B_q=1\gev^{-2}$). The  $|t|$ slopes extracted from the IPsat calculation of the incoherent cross section at two different Bjorken-$x$ values are shown in Fig.~\ref{fig:deuteron_incoh_slope_ipsat}. As the size of the nucleons and hot spots does not depend on $x$, the extracted slopes are independent of $x$.

In the CGC calculation, on the other hand, both nucleons and hot spots grow as a result of the small-$x$ evolution. The extracted slopes in this case are shown in Fig.~\ref{fig:deuteron_incoh_slope}. At moderate $|t|\sim 0.2\gev^2$ the result of the $x$ evolution is to make the spectra steeper, because of the growth of the system, both in the case with and without nucleon shape fluctuations. For any $x$, at large $|t|$ nucleon shape fluctuations start to dominate if included, similar to the IPsat calculation. Without nucleon shape fluctuations, the slopes are not constant at large $|t|$ in contrast to the IPsat calculation. This is due to the short-range color charge fluctuations in the target that contribute to the variance of the scattering amplitude.
As a result of the small-$x$ evolution, these fluctuations start to take place at shorter distance scales, and consequently the cross section falls more slowly as a function of $|t|$ after the evolution. Because of this, the slope as a function of $|t|$ for smaller $x$ crosses that for the initial $x$ at an intermediate $|t|\approx 0.6\gev^2$.

We note that, in case of an infinite target, in the limit $|t| \gg Q_s^2$ and for small dipoles, the incoherent cross section approaches a power-law, which dominates over the exponentially falling geometric component at large enough $|t|$. When the nucleon shape fluctuations are included, the short-range color charge fluctuations do not have a large effect as the results are very similar to the case of the IPsat calculation in the studied $|t|$ range.

\subsection{Evolution of the Helium structure}

\begin{figure}[tb]
\centering
		\includegraphics[width=0.48\textwidth]{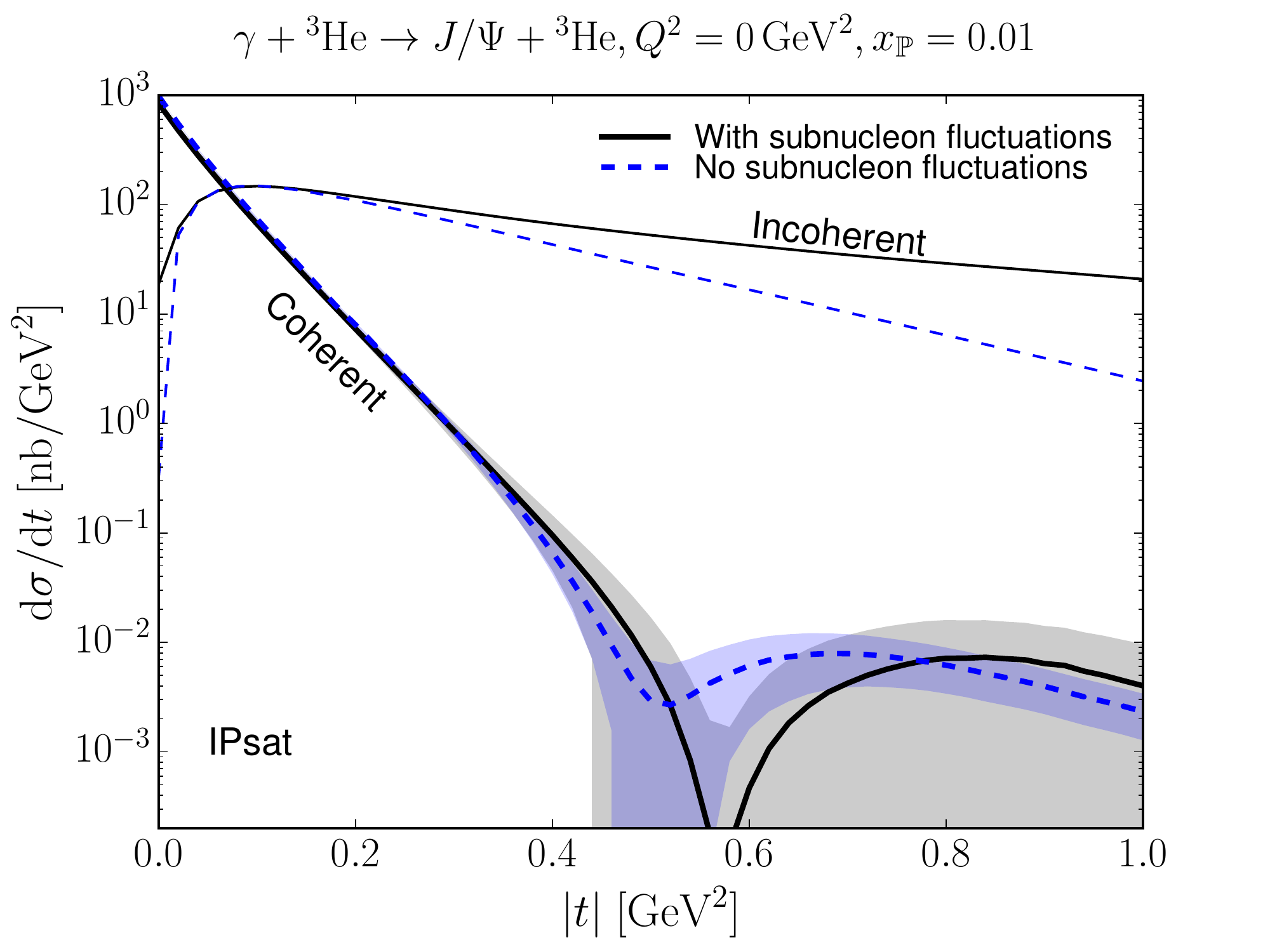} 
				\caption{Coherent and incoherent diffractive $J/\Psi$ photoproduction cross section in photon-$^3$He collisions at $\xpom=0.01$ with (solid lines) and without (dashed lines) subnucleon fluctuations, calculated using the IPsat parametrization. }
		\label{fig:helium_spectra}
\end{figure}

Similar to the case of deuterons, we study the effect of nucleon shape fluctuations in ${}^3$He by calculating diffractive $J/\Psi$ photoproduction cross sections off helium-3 in the EIC kinematics, first using the IPsat parametrization to describe dipole-nucleon scattering. The results are shown in Fig.~\ref{fig:helium_spectra}, where we again find that the effect of nucleon shape fluctuations changes the incoherent cross section at $|t|\gtrsim 0.2\gev^2$, corresponding to the distance scale of the nucleon substructure used in our calculations.

\begin{figure}[tb]
\centering
		\includegraphics[width=0.48\textwidth]{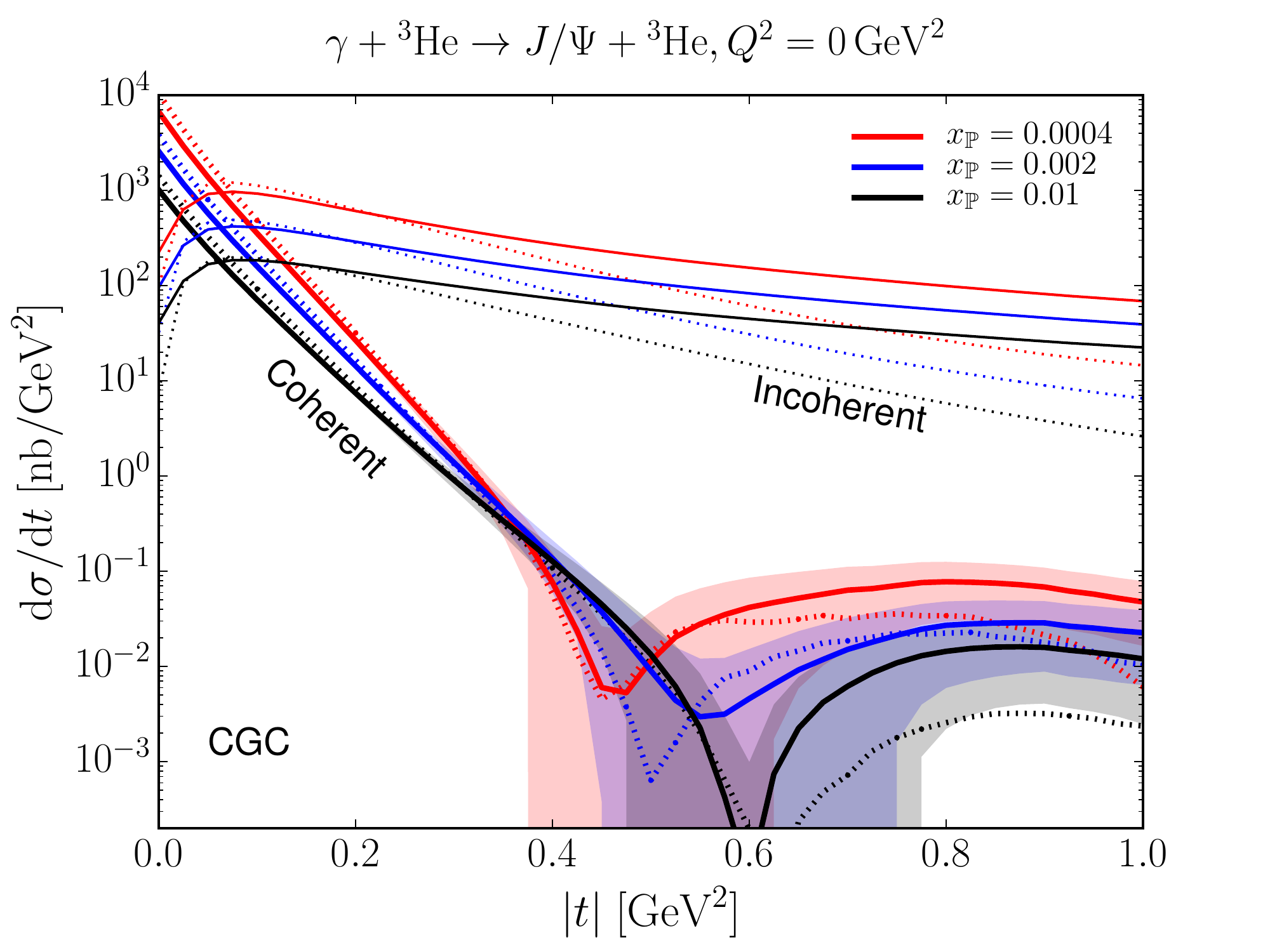} 
				\caption{Coherent and incoherent diffractive $J/\Psi$ photoproduction cross sections in photon-$^3$He collisions at different $\xpom$.  Solid lines include nucleon shape fluctuations and dotted lines do not. For clarity, statistical uncertainty of the calculation is only shown for the case with fluctuating substructure, where errors are much larger. }
		\label{fig:helium_spectra_cgctl}
\end{figure}

To study the energy evolution of the helium structure illustrated in Fig.~\ref{fig:helium_density}, we calculate coherent and incoherent $J/\Psi$ production cross sections at the initial condition and at smaller $x$ values accessible at the EIC. Similar to deuterons, the position of the first diffractive minimum moves to smaller $|t|$ as a result of the evolution, and the coherent spectra get steeper at small $|t|$. Incoherent cross sections in both cases are comparable up to $|t|\sim 0.2\gev^2$, after which  the spectra are much more steeply falling in case of no nucleon substructure fluctuations.

\section{Conclusions and Outlook}\label{sec:conclusions}

We presented predictions for exclusive $J/\Psi$ photoproduction in the EIC kinematics with deuteron and Helium-3 targets. We showed that employing two commonly used  deuteron  wave functions  resulted in significantly different coherent vector meson spectra. This demonstrates the sensitivity for probing the transverse distribution of small-$x$ gluons at a future EIC. In particular, diffractive vector meson measurements at an EIC could reveal deviations in the spatial gluon distributions of light nuclei from the model assumptions in this work.

By solving the small-$x$ JIMWLK evolution equation, we predicted the energy dependence of the coherent and incoherent cross sections in the EIC energy range. The slope of the coherent $|t|$-dependent cross sections was found to become steeper due to the growth of the nucleus with decreasing $x$. The differences between the coherent cross sections computed with different wave functions at $\xpom=0.01$ remained similar down to $\xpom$ of a few times $10^{-4}$.

We showed that the incoherent cross section at $|t|\gtrsim 0.2\gev^2$ is sensitive to additional nucleon substructure fluctuations, which were previously constrained by HERA data. These results are similar to what was found for heavy nuclei~\cite{Mantysaari:2017dwh}. The small-$x$ evolution did not significantly modify the $|t|$ value above which the nucleon substructure affects the incoherent cross section. However, the $|t|$ slope of the incoherent cross section also becomes steeper with decreasing $x$, which indicates the growth of the fluctuating constituents, namely both nucleons ($|t| \sim 0.2\gev^2$) and subnucleonic hot spots ($|t| \gtrsim 0.2\gev^2$).

When not considering geometric subnucleon fluctuations, the slope of the incoherent vector meson spectra at $|t|\gtrsim 0.6\gev^2$, becomes flatter as a result of the evolution. This is because with increasing $Q_s$, color charge fluctuations appear on decreasing distance scales. This effect is absent in calculations without explicit color charge fluctuations, as shown in case of the IPsat model.

Besides the fundamental information on the gluonic structure of light nuclei at small $x$ that exclusive vector meson production can provide, it also has applications for the phenomenology of high energy nucleus-nucleus collisions. 
To interpret deuteron-gold and helium-gold measurements at RHIC, it is important to have precise knowledge of the small-$x$ geometry of light ions, which is an input for model calculations involving hydrodynamic simulations of QGP evolution~(see \cite{Dusling:2015gta} for a review and \cite{Romatschke:2015gxa,Shen:2016zpp,Weller:2017tsr,Mantysaari:2017cni,Schenke:2019pmk} for more recent developments). 

 Systematically going to heavier nuclei  will be very interesting, as clustering of nucleons and other correlations, and their effect on the small $x$ gluonic structure could be probed by a future EIC as well (see also~\cite{Kowalski:2007rw,Kowalski:2008sa,Mantysaari:2017slo} for a discussion of systematics of saturation effects in heavier nuclei).

\section*{Acknowledgments}
We thank T. Haverinen, M. Mace, F. Salazar and T. Ullrich for discussions. B.P.S. is supported under DOE Contract No. DE-SC0012704. H.M. is supported by the Academy of Finland, project 314764, and by the European Research Council, Grant ERC-2015-CoG-681707. H.M wishes to thank the Nuclear Theory Group at BNL for hospitality during the final stages of this work.
 This research used resources of the National Energy Research Scientific Computing Center, which is supported by the Office of Science of the U.S. Department of Energy under Contract No. DE-AC02-05CH11231, CSC -- IT Center for Science in Espoo, Finland and  the Finnish Grid and Cloud Infrastructure (persistent identifier \texttt{urn:nbn:fi:research-infras-2016072533}).

\bibliographystyle{JHEP-2modlong.bst}
\bibliography{refs}

\pagebreak

\end{document}